\documentclass[onecolumn,aps,prb, superscriptaddress]{revtex4-2}
\usepackage{epsfig}
\usepackage{graphicx}
\usepackage{color}
\usepackage{amsmath}
\usepackage{amsfonts}
\usepackage{enumitem}
\usepackage{hyperref}
\hypersetup{
	colorlinks = true,
	linkcolor = blue,
	citecolor = blue
}
\newcommand{\be}{\begin{equation}}
\newcommand{\ee}{\end{equation}}

\newcommand{\bea}{\begin{eqnarray}}
\newcommand{\eea}{\end{eqnarray}}

\newcommand{\p}{\partial}

\newcommand{\la}{\left\langle}
\newcommand{\ra}{\right\rangle}

\renewcommand{\vec}[1]{{\bf #1}}
\renewcommand{\phi}{\varphi}
\renewcommand{\epsilon}{\varepsilon}
\def\nn{\nonumber\\}

\renewcommand{\cite}[1]{[\onlinecite{#1}]}

\begin{document}

\title{Current distribution and group velocities for electronic states on $\alpha-\mathcal{T}_3$ lattice ribbons in a magnetic field}
\date{\today}

\author{D. O. Oriekhov}
\affiliation{Instituut-Lorentz, Universiteit Leiden, P.O. Box 9506, 2300 RA Leiden, The Netherlands}

\author{Sergey Voronov}
\affiliation{National Technical University of Ukraine “Kiev Polytechnic Institute”, Peremogy Ave. 37, 03056, Kiev, Ukraine}

\begin{abstract}

 We study the group velocities of electronic states and distributions of currents in $\alpha-\mathcal{T}_3$ lattice ribbons under a uniform perpendicular
 magnetic field. Using the effective low-energy model we analyze all possible simple configurations of lattice termination with zigzag and armchair boundaries. 
 We show that the edge current depends on the type of zigzag termination, and can be zero or finite near the edge. Also similar dependence is observed in the case of armchair termination and is related to the size of the ribbon. The nonzero current flowing along the edge can be used a signature of formation of propagating edge states. Also we show the qualitative difference in the distribution of the edge current between the case of $\alpha=1$ (dice model) and other values of model parameter $\alpha\neq 1$ for armchair-terminated ribbons.
\end{abstract}
\maketitle

\section{Introduction}

Recently it was discovered [\onlinecite{Bradlyn}] that in crystals with special space and point symmetry groups a sophisticated electron spectra with high pseudospins could be realized with no analogues
in particle physics. For example, the spectra of corresponding quasiparticles could
possess strictly flat bands [\onlinecite{Heikkila1,Heikkila2,Leykam}]). The
dice model is the paradigmatic and historically the first example of such a system with a flat band which hosts pseudospin-1 fermions [\onlinecite{Sutherland}]. In this paper we study its' generalization, which is called $\alpha-\mathcal{T}_3$ model \cite{Raoux}. This is a tight-binding model of two-dimensional fermions living on the so-called ${\cal T}_3$ (or dice) lattice where atoms are situated both at the vertices
of hexagonal lattice and the hexagons centers [\onlinecite{Sutherland,Vidal}]. The $\alpha-\mathcal{T}_3$ model has three sites per unit cell and the electron states in
this model are described by three-component fermions. The corresponding quasiparticle energy spectrum is comprised of three bands, the two of them are Dirac cones and the third one is completely flat and has zero energy [\onlinecite{Raoux}]. All three bands meet at the $K$ and $K^{\prime}$ points, which are situated at the corners of the Brillouin zone. The ${\cal T}_3$ lattice has been experimentally realized in Josephson arrays [\onlinecite{Serret,Abilio1999}] and metallic wire networks [\onlinecite{Naud}], and the possible optical realization was proposed in Ref.[\onlinecite{Rizzi}].

The properties of $\alpha-\mathcal{T}_3$ were extensively studied in recent years \cite{Raoux, Malcolm2016, Nicol, Biswas2016}. For example, its magnetotransport properties, such as the
collisional and Hall conductivity for the $\alpha-\mathcal{T}_3$ model were calculated in Refs.\cite{Biswas2016}. In this paper we concentrate our attention on the properties of $\alpha-\mathcal{T}_3$ model on terminated lattice placed in perpendicular magnetic field. Such geometry is typical for the recent  magnetotransport experiments on graphene ribbons \cite{Richter2020Nature}. 

The properties of ribbons made of $\mathcal{T}_3$ lattice were already studied in Refs.\cite{Xu2017PRB,Oriekhov2018FNT,Bugaiko2019JPCM,Yan-Ru-Chen2019, Dey2019PRB,Dey2020PRB-Haldane,Soni2020PRB}. In a recent paper \cite{Oriekhov2018FNT}, the classification of possible termination and low-energy boundary conditions was given for dice lattice. In the Ref.\cite{Bugaiko2019JPCM} the above classification were used to study the spectral properties of $\mathcal{T}_3$ lattice ribbons in perpendicular external magnetic field. We apply the results of these papers to analyze the group velocities and current distributions for the lowest Landau levels. The optical response of $\mathcal{T}_3$ lattice ribbons with all combinations of simple termination
was addressed in the Ref. \cite{Yan-Ru-Chen2019}. In the recent paper \cite{Dey2019PRB} the radiation-dressed band structure of $T_3$ lattice ribbon with armchair edges was analyzed, with the emphasize on the appearance of propagating edge states. The formation of edge state in the Haldane model for dice nanoribbons was discussed in Ref.\cite{Dey2020PRB-Haldane}. The existence of flat bands near the Fermi level, edge
currents and edge charge localization near zero energy for open boundary conditions was shown for dice lattice ribbons with Rashba spin-orbit coupling placed in magnetic field \cite{Soni2020PRB}. Interestingly, for dice lattice ribbons with additional $S_z$ mass the appearance of in-gap edge states, which are degenerate in zigzag termination case, was predicted in Ref.\cite{Xu2017PRB} 

These results motivate us to study the distribution of currents for the dice lattice ribbons placed in magnetic field. 
A similar study in the case of graphene semi-infinite lattices \cite{Wang2011} uncovered peculiar properties of each termination type, for example, the large edge current densities near armchair edges and the universal behavior of total currents. As was shown in Refs.\cite{Oriekhov2018FNT, Bugaiko2019JPCM} there are no edge states for any termination in dice $(\alpha=1)$ model, but they are formed for the $\alpha\neq 1$ parameter values or due to the presence of magnetic field. In the present paper we analyze their manifestation in current distributions.  
Also, one should note that the flat band does not play role in the formation of currents distribution since it consists of localized states \cite{Raoux,Bercioux, Gorbar2019PRB, Oriekhov2018FNT}. For the zigzag terminated case we show that the presence of edge current strongly depends on the termination type, and completely vanishes in several cases. In armchair case we find that while the spectrum of the ribbons was qualitatively the same for any width \cite{Bugaiko2019JPCM}, the distribution of currents is drastically different for the so-called "metallic" and "armchair" numbers of atomic rows in the ribbon. Together these results can be used as a clear signature of the presence of absence of edge states with nonzero energy in a terminated $\mathcal{T}_3$ lattice.

The paper is organized as follows. In Sec. \ref{sec:model} we
first discuss the $\alpha-\mathcal{T}_3$ model and recall the main termination types and boundary conditions. Next, in Sec.\ref{sec:general} we proceed to the $\mathcal{T}_3$ lattice ribbons with zigzag termination, that is infinite in one direction. We analyze the dependence of group velocities and edge currents on Landau level index, and show that in zigzag termination case the behavior of these quantities differs from the graphene case. We analyze all different combinations of zigzag terminations and show how the current distribution depends on the termination type. Also the group velocity is studied for semi-infinite lattice to gain an intuitive insight into the role of each edge. Next, in Sec.
\ref{sec:armchair} we analyze the armchair termination, and discuss the influence of magnetic field and ribbon width on currents distribution. Finally, we discuss the main results and conclusions in Sec.\ref{sec:conclusion} 

\section{The $\alpha-\mathcal{T}_3$ model}
\label{sec:model}
In this section we review the main properties of free $\alpha-\mathcal{T}_3$ model and the classification of simple lattice terminations and corresponding boundary conditions.
\begin{figure}
	$\text{(a)}$\includegraphics[scale=0.33]{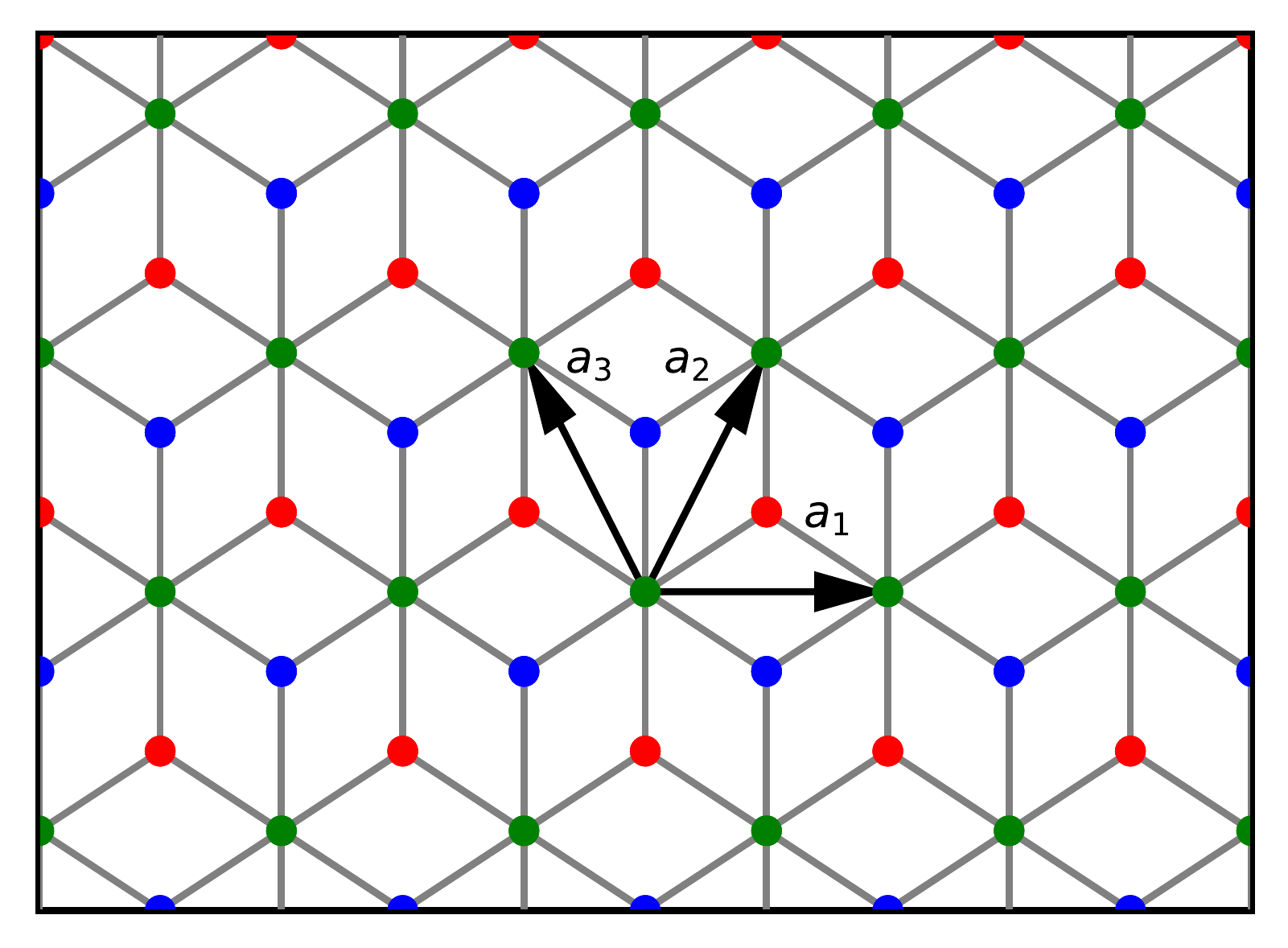}
	$\text{(b)}$\includegraphics[scale=0.33]{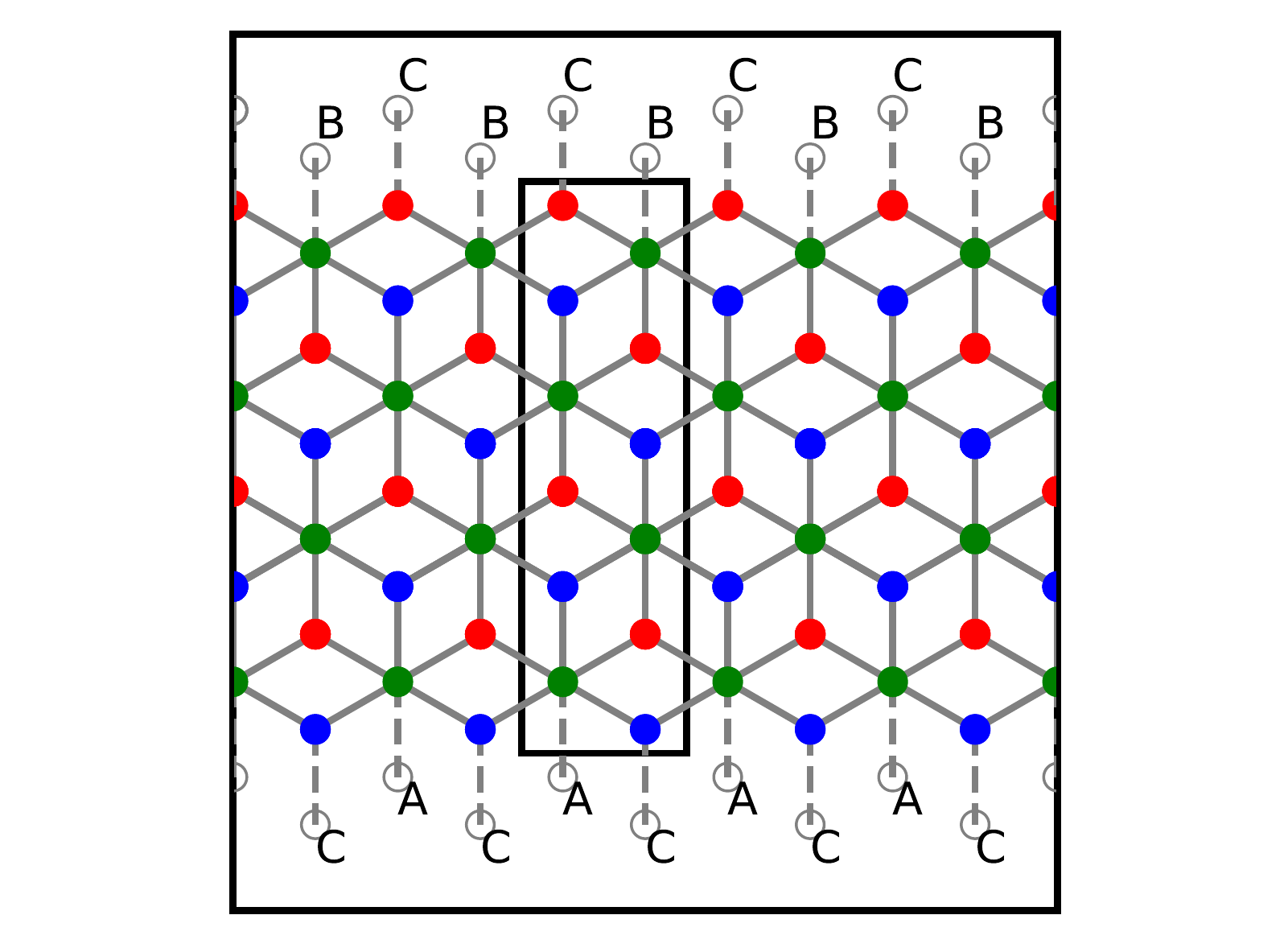}
	$\text{(c)}$\includegraphics[scale=0.33]{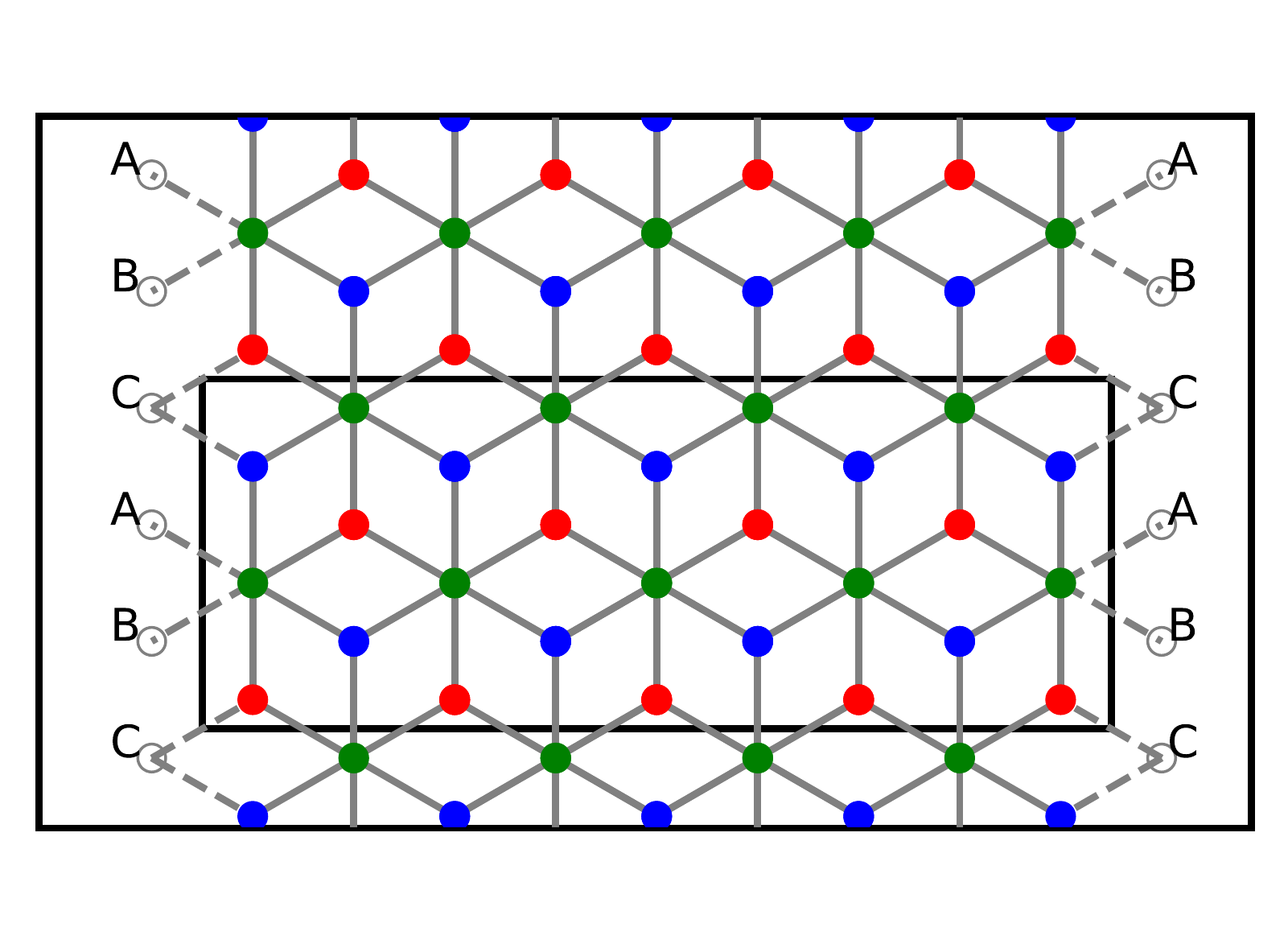}
	\caption{The ${\cal T}_3$ lattice whose red points display the atoms of the
		$A$ sublattice, the blue points describe the $B$ sublattice, and the green points define the $C$ sublattice. On the panel (a) the vectors
		$\vec{a}_1=(\sqrt{3},\,0)d$ and $\vec{a}_2=(\sqrt{3}/2,\,3/2)d$ are the basis vectors of the $C$ sublattice. Panel (b) shows zigzag termination example, and panel (c) demonstrates the armchair one. Black rectangle measures the actual ribbon width $L$ in units of lattice constant.}
	\label{fig1}
\end{figure}

The $\alpha-\mathcal{T}_3$ model describes quasiparticles in two dimensions with pseudospin $S=1$ on the so-called dice lattice schematically shown in
Fig.\ref{fig1} \cite{Raoux}. This lattice has a unit cell with three different lattice sites whose two sites ($A,C$) like in graphene form a honeycomb
lattice with hopping amplitude $t_{AC}=t_1$ and additional $B$ sites at the center of each hexagon are connected to the $C$ sites with
hopping amplitude $t_{BC}=t_2$. The two hopping parameters $t_1$ and $t_2$ are not equal, in general, and the dice model corresponds to the limit $t_1=t_2$. The Brillouin zone of this lattice is the same as for graphene because the underlying sublattices $A$, $B$ and $C$ are triangular Bravais lattices (see Ref.\cite{Katsnelson} for the discussion of graphene case). The local topology of couplings on dice lattice \cite{Leykam} protects the flat band against perturbations. The bipartite symmetry is present because the `hub' sites ($C$) are coupled only to `rim' sites ($A,\,B$) and vise versa. The tight-binding Hamiltonian of the model in momentum space reads \cite{Raoux}
\begin{align}
	\label{TB-Hamiltonian}
	H=\left(\begin{array}{ccc}
		0 & f_{\vec{k}}\cos\Theta & 0\\
		f^{*}_{\vec{k}}\cos\Theta & 0 & f_{\vec{k}}\sin\Theta\\
		0 & f^{*}_{\vec{k}}\sin\Theta & 0
	\end{array}\right),
	\quad \alpha \equiv \tan\Theta=\frac{t_2}{t_1},\quad f_{\vec{k}}=-\sqrt{t_1^2+t_2^2}\,(1+e^{-i\vec{k}\vec{a}_2}+e^{-i\vec{k}\vec{a}_{3}}).
\end{align}
Here $\vec{a}_1=(\sqrt{3},\,0)d$ and $\vec{a}_2=(\sqrt{3}/2,\,3/2)d$ are the basis vectors of triangle $C$ sublattice. The basis vectors of corresponding reciprocal lattice are $\mathbf{a}_{1}^{*}=2 \pi/\sqrt{3} d \left(1,-1/\sqrt{3}\right)$ and $\quad \mathbf{a}_{2}^{*}=(0, 4 \pi/3 d)$. They are shown together with the lattice in Fig.\ref{fig1}, and $d$ denotes the nearest-neighbor interatomic distance. The energy spectrum of the above Hamiltonian is qualitatively the same for any $\alpha$ and consists of
three bands: the zero-energy flat band, $\epsilon_0(\mathbf{k})=0$, and two dispersive bands $\epsilon_{\pm}(\mathbf{k})=\pm|f_{k}|$.
The six values of momenta, for which $f_{\vec{k}}=0$, correspond to the three bands touching points and called $K$ points. They are situated at the corners of the hexagonal
Brillouin zone. One can select the two non-equivalent points as
\begin{align}
	\vec{K}=\frac{2\pi}{d}\left(\frac{\sqrt{3}}{9},\,\frac{1}{3}\right),\quad \vec{K}'=\frac{2\pi}{d}\left(-\frac{\sqrt{3}}{9},\,
	\frac{1}{3}\right).
\end{align}
The four remaining corners of Brillouin zone may be connected to one of these points via a translation by a reciprocal lattice vector.
For momenta near the $K$-points, $\vec{k}=\vec{K}(\vec{K}')+\tilde{\vec{k}}$, we find that $f_{\vec{k}}$ is linear in $\tilde{\vec{k}}$,
i.e., $f_{\vec{k}}=\hbar v_F(\lambda \tilde{k}_x-i\tilde{k}_y)$ with valley index $\lambda=\pm$, where $v_F=3td/2\hbar$ is the Fermi velocity, and in what follows
we omit for the simplicity of notation the tilde over momentum. Thus, we obtain the low-energy Hamiltonian near the $K(K^{\prime})$-point in the
form \cite{Malcolm2016}
\begin{align}\label{Hd-hamiltonian}
	&\mathcal{H}_{\lambda}=\hbar v_F(\lambda S_x k_x+S_yk_y)=\hbar v_F\left(\begin{array}{ccc}
		0 & \cos\Theta (\lambda k_{x}-ik_y) & 0\\
		\cos\Theta (\lambda k_{x}+ik_y) & 0 & \sin\Theta  (\lambda k_{x}-ik_y)\\
		0 & \sin\Theta (\lambda k_{x}+ik_y) & 0
	\end{array}\right), \nn
	& S_x=\left(\begin{array}{ccc}
		0 & \cos\Theta & 0\\
		\cos\Theta & 0 & \sin\Theta\\
		0 & \sin\Theta & 0
	\end{array}\right), \quad S_y=\left(\begin{array}{ccc}
		0 & -i\cos\Theta & 0\\
		i\cos\Theta & 0 & -i\sin\Theta\\
		0 & i\sin\Theta & 0
	\end{array}\right),
\end{align}
where $\mathbf{S}$ are the spin matrices of the spin 1 representation. The Hamiltonian acts on
three-component wave functions $\Psi^T=(\Psi_{A},\Psi_{C},\Psi_{B})$. The full Hamiltonian, which includes both valleys, is  given by block-diagonal matrix $\text{diag}(H_{+},H_{-})$ and acts on 6-component spinors $(\Psi_{+},\Psi_{-})^{T}$.

It is straightforward to describe the interaction with a magnetic field via the standard Peierls substitution
$\vec{k}\to \vec{k}+\frac{e}{\hbar c}\vec{A}$ in the Hamiltonian. In the following we will use the freedom of the choice of the gauge of vector
potential $\vec{A}$ in order to simplify calculations in particular geometries.

The boundary conditions are determined from the condition that the matrix element of electric current normal to the boundary vanishes, $\la\Psi_B\right|(\vec{J}_{+}+\vec{J}_{-})\vec{n}\left|\Psi_B\right\rangle=0$. Here the current operator is defined as $ \vec{n}\vec{J}_{\lambda}=\lambda S_x n_x+S_y n_y$, and the index $\lambda=\pm$ stands for the valley $K(K')$. This particular form of the current operators follows from the low-energy Hamiltonian \eqref{Hd-hamiltonian}, which is linear in momentum. Thus the current operator is not a differential operator. This property of current operator is usually used to derive the proper boundary conditions in systems with Dirac dispersion \cite{Oriekhov2018FNT,Urban2011,Akhmerov}. The simple types of termination can be classified in the same way as in graphene \cite{Akhmerov} - into zigzag (usually along $x$-direction) and armchair (along $y$-direction) types. At the same time, the zigzag termination type supplies much more rich variety of boundary conditions for the dice lattice \cite{Oriekhov2018FNT} than in graphene. The example of ribbons with both types of termination are presented on Fig.\ref{fig1}.

Next we proceed to the detailed discussion of group velocities and current distributions in each termination case. Since the zigzag termination demonstrates many new properties comparing to graphene \cite{Oriekhov2018FNT, Bugaiko2019JPCM} (while armchair termination is quiet similar to graphene case), we mainly concentrate the attention on zigzag ribbons. 

\section{Lattice infinite in x-direction: zigzag termination}
\label{sec:general}
\begin{figure}
	$\text{(a)}$\includegraphics[scale=0.47]{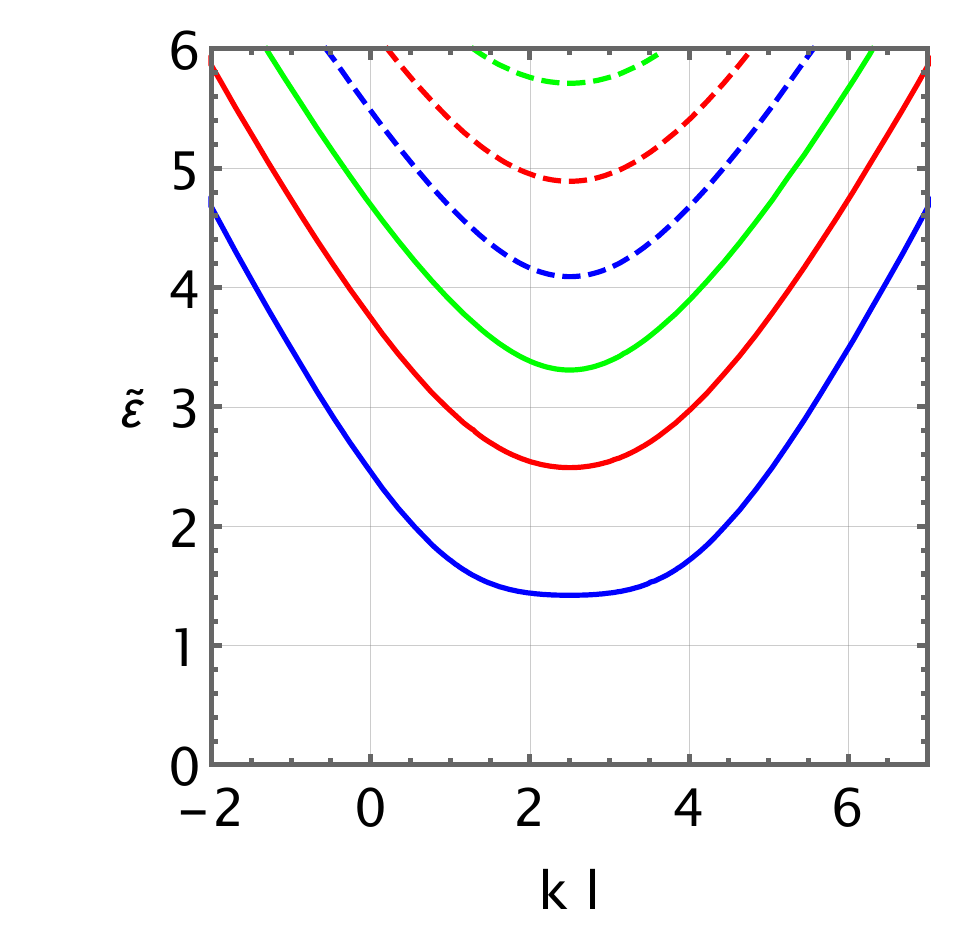}\qquad
	$\text{(b)}$\includegraphics[scale=0.47]{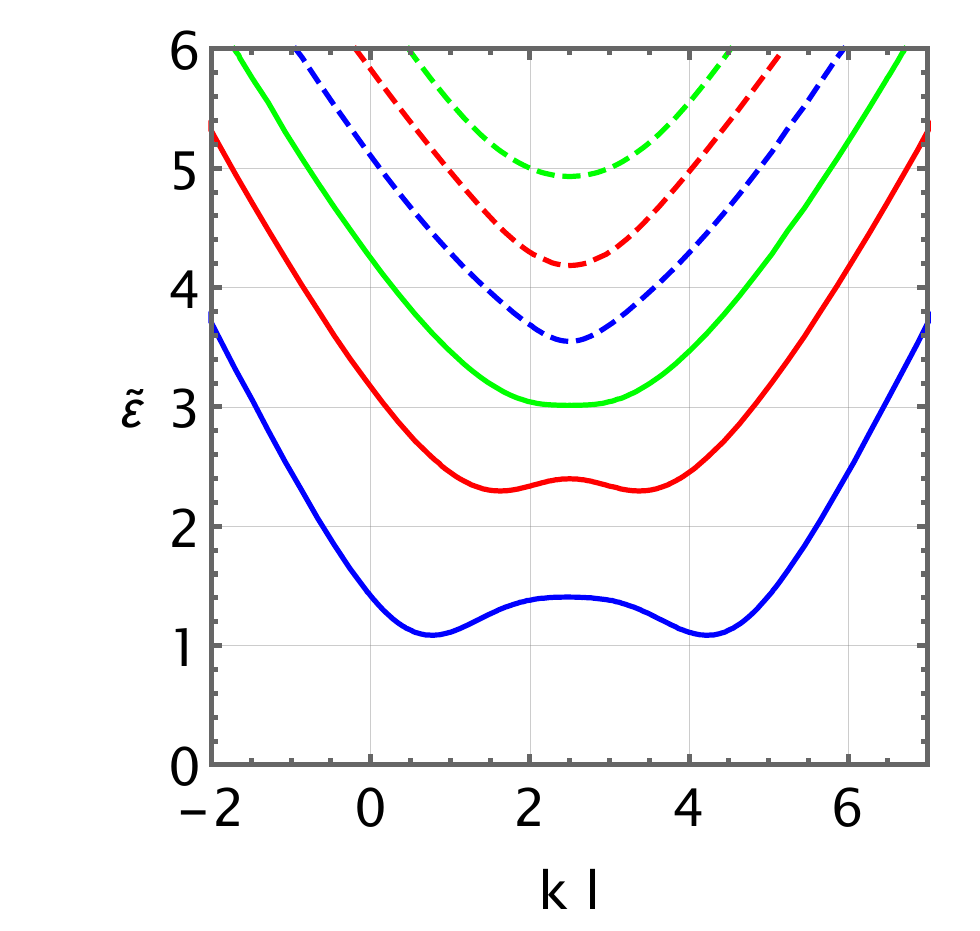}\qquad
	$\text{(c)}$\includegraphics[scale=0.47]{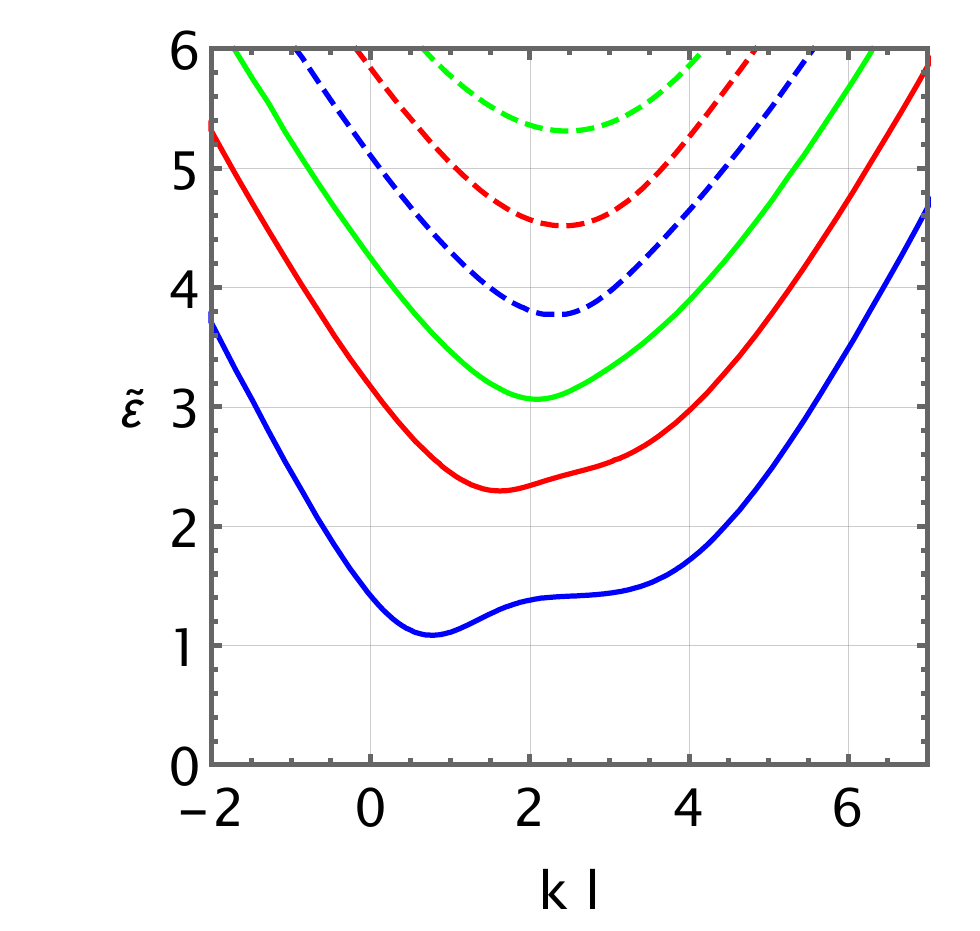}
	\caption{Spectrum of zigzag terminated ribbons $\tilde{\epsilon}$ as a function of wave number $k\equiv k_x$ for the width $L=5 l$. The value of parametric angle is $\Theta=\frac{\pi}{4}$ and corresponds to the dice model case. Termination types: (a) - C-C, (b) - AB-BA, (c) - AB-C. Note that on panels (a) and (b) the spectrum is symmetric with respect to central wave number $k l=k_0 l /2=2.5$.}
	\label{fig:spectrum}.
\end{figure}
For the ribbon shown on panel (b) in Fig.\ref{fig1}) we choose the gauge in the form $A = (-By, 0)$, which preserves
translational invariance in the $x$ direction. Then the wave functions can be chosen in the form
$\Psi_\mu = e^{i k_x x}\psi_\mu$, and the Schr\"{o}dinger equation becomes
\begin{align}\label{eq:free_system}
	\begin{pmatrix}
		0 & \cos{\Theta} (\lambda \xi +\p_\xi) & 0 \\
		\cos\Theta(\lambda \xi -\p_\xi) & 0 & \sin\Theta(\lambda \xi +\p_\xi) \\
		0 & \sin\Theta(\lambda \xi -\p_\xi) & 0
	\end{pmatrix}
	\begin{pmatrix}
		\psi_A \\
		\psi_C \\
		\psi_B
	\end{pmatrix}
	= \frac{\widetilde{\epsilon}}{\sqrt{2}}
	\begin{pmatrix}
		\psi_A \\
		\psi_C \\
		\psi_B
	\end{pmatrix},\quad \xi = k_x l - y/l.
\end{align}
Here we are working in the notation from Ref.\cite{Bugaiko2019JPCM} $\widetilde{\epsilon} = \frac{2\epsilon}{\epsilon_0}$, where  $ l = \sqrt{\hbar c/|eB|}$ is the magnetic length, and
$\epsilon_0 = \sqrt{2\hbar v_F^2 |eB|/c}$ is Landau energy scale. 
The first and third lines of the system define $\psi_A$ and $\psi_B$ in terms of $\psi_C$ in the case $\widetilde{\epsilon}\neq 0$
\begin{align}
	\psi_{A} = \sqrt{2}\cos\Theta \frac{\lambda\xi + \partial_\xi}{\widetilde{\epsilon}} \psi_{C},\quad
	\psi_{B} = \sqrt{2}\sin\Theta \frac{\lambda\xi - \partial_\xi}{\widetilde{\epsilon}} \psi_{C}.
\end{align}
The second line of system \eqref{eq:free_system} gives the second-order differential equation for $\psi_{C}$:
\begin{align}\label{eq:eq_on_psi_C}
	(\partial^2_\xi - \xi^2)\psi_C + \left(\lambda \cos 2\Theta + \frac{\widetilde{\epsilon}^2}{2}\right)\psi_C = 0,
\end{align}
which solution can be expressed in terms of the parabolic cylinder functions $U$ and $V$ \cite{Abramowitz}
\begin{align}\label{eq:zigzag_general}
	\psi_C(y) = C_1U\left(-\frac{\widetilde{\epsilon}^2}{4} - \frac{\lambda \cos 2\Theta}{2},\sqrt{2}\xi\right) + C_2V\left(-\frac{\widetilde{\epsilon}^2}{4} - \frac{\lambda \cos 2\Theta}{2},\sqrt{2}\xi\right),
\end{align}
where $C_1$ and $C_2$ are arbitrary constants. From this solution one can find the spectrum of infinite system $
	\epsilon_{n}(\Theta) = \pm \epsilon_0\sqrt{n + 1/2(1 - \lambda \cos 2 \Theta)}$ 
\cite{Bercioux,Raoux,Bugaiko2019JPCM}.
which is different in the $K$ and $K'$ valleys for $\Theta\neq 0,\frac{\pi}{4}$. 

In our case we need to plug these solutions into the boundary conditions at ribbon edges $y=0$ and $y=L$ to determine the energy spectrum and constant $C_1$ and $C_2$. Since the calculations of dispersion were discussed in great detail in Ref.\cite{Bugaiko2019JPCM}, below we will mainly focus on the evaluation of group velocity and current. 

Firstly, let us recall the main types of zigzag boundary conditions. 
From the requirement of vanishing of the normal current at the boundary one finds the following restriction on wave functions:
\begin{align}
	\psi_C\bigg|_{B} = 0,\quad \text{and} \quad (\psi_A \cos\Theta - \psi_B \sin\Theta)\bigg|_{B} = 0.
	\end{align}
 As was found in
	\cite{Oriekhov2018FNT}, the first condition (below we called it "C" condition) corresponds to the C, AC or BC types of lattice termination at low energies, while the last one corresponds
	to the AB termination (and called "AB"). By using Schrodinger equation, one can rewrite the second condition as 
	$\psi_C' + \xi \cos2\Theta\psi_C = 0$. Thus, one finds 3 main combinations of boundary conditions: C-C, BA-C, and BA-AB. Next, we present the calculation on example of C-C boundary conditions and discuss the physical results for other combinations, leaving technical details in the Appendix \ref{appendix}.

\subsection{Group velocity}
\begin{figure}
	\hspace{-0.5cm}$\text{(a)}$\includegraphics[scale=0.47]{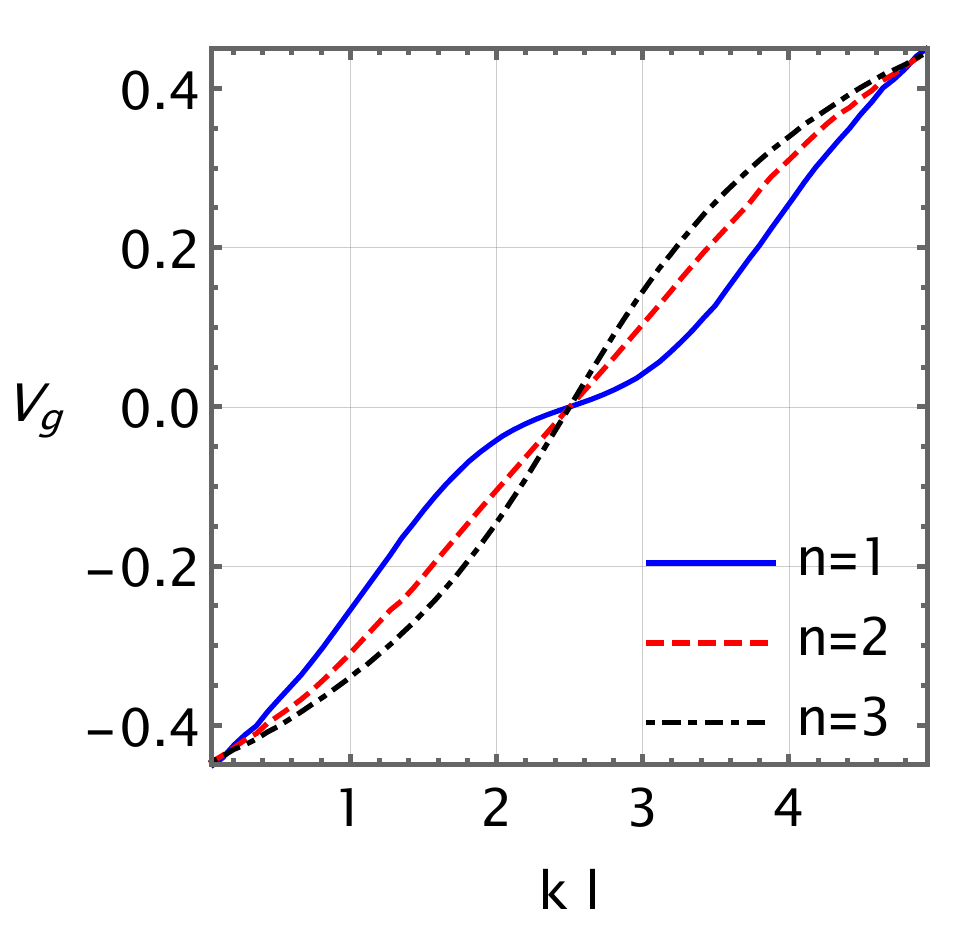}\qquad
	$\text{(b)}$\includegraphics[scale=0.47]{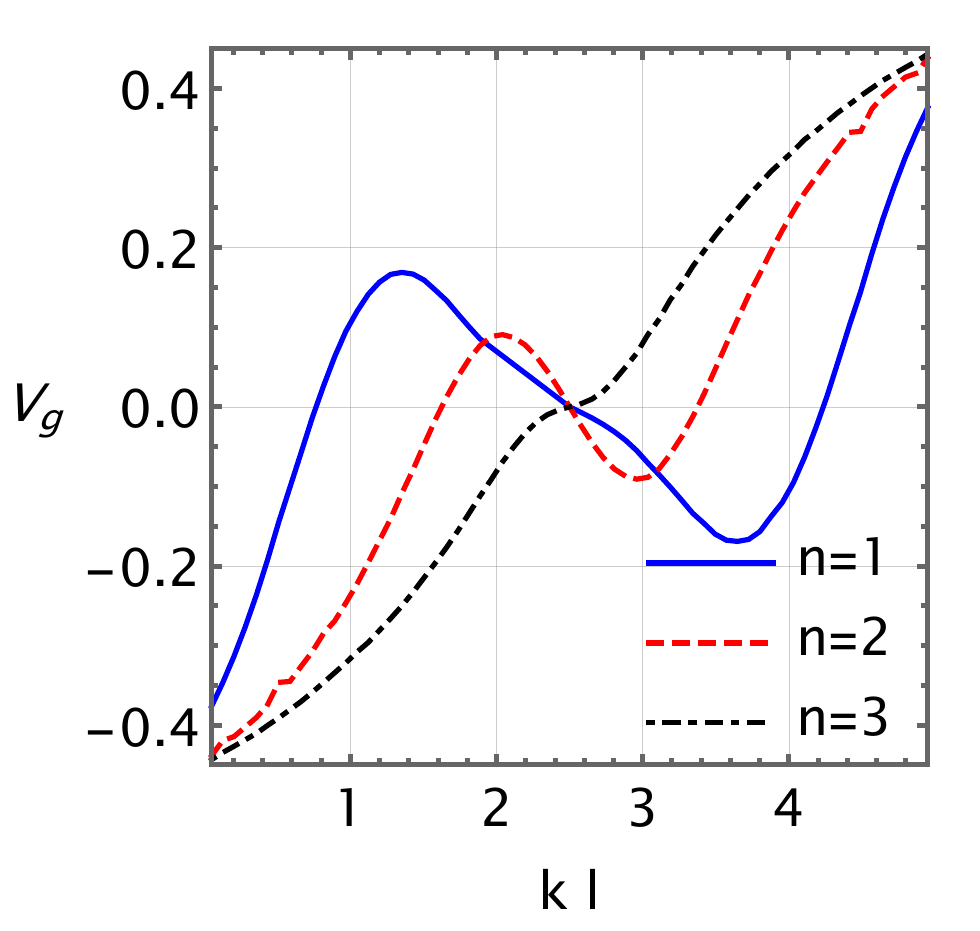}\qquad
	$\text{(c)}$\includegraphics[scale=0.47]{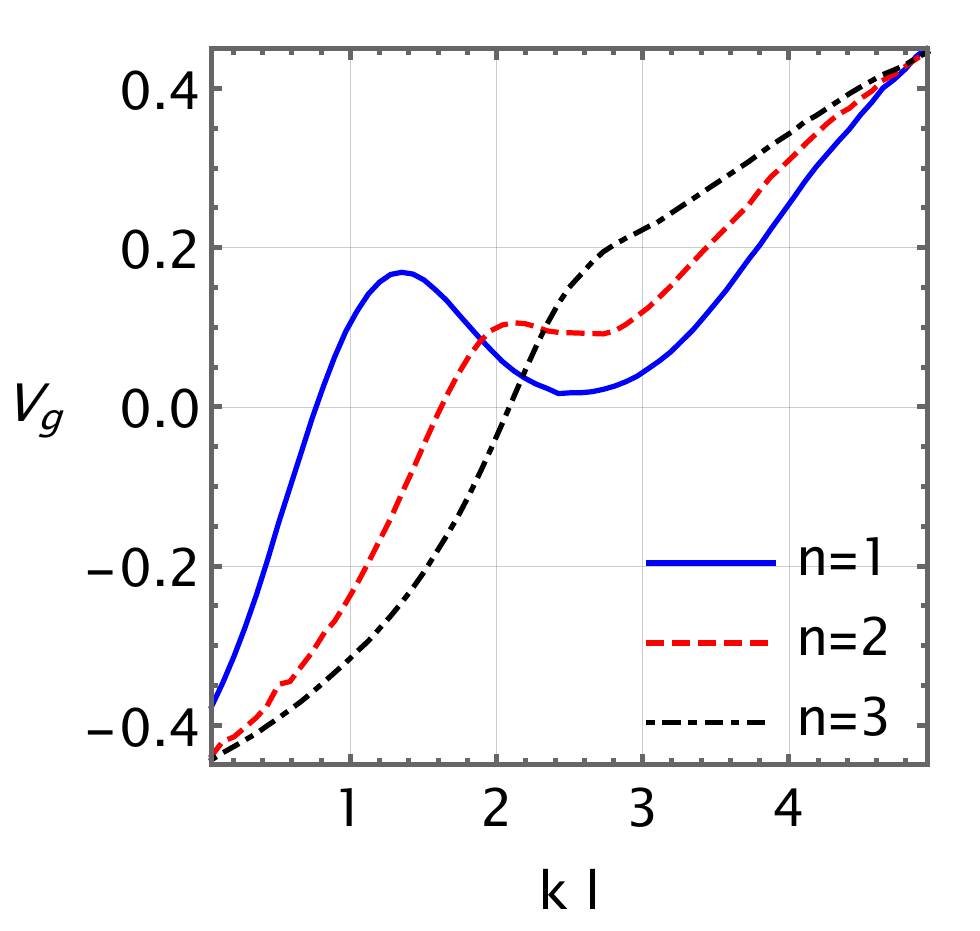}
	\caption{Group velocities $V_{g}^{(x)}$ for the first three Landau levels for zigzag terminated ribbons as a functions of wave number $k\equiv k_x$. Termination types: (a) - C-C, (b) - BA-AB, (c) - BA-C. The width of the ribbon is $L=5 l$ and $\Theta=\frac{\pi}{4}$.}
	\label{fig:group-vel-zigzag}
\end{figure}
The group velocity is defined for $n$-th energy level $E_{n}(k_\mu)$ as $V_g^{(\mu)}(n,\vec{k})=\frac{1}{\hbar} \frac{\p E_{n}(k_\mu)}{\p k_\mu}.$ If the spectral equation is expressed in terms of special functions such as parabolic cylinder functions, it is not always possible to find the exact analytic dependence $E_{n}(k_\mu)$. In such a case one can find the expression for group velocity from the dispersion equation itself. Suppose the equation is $f(E_n(\vec{k}),\vec{k})=0$. Then, by differentiating the equation with respect to $k_\mu$, we find
\begin{align}\label{eq:group-velocity}
	\hbar V_g^{(\mu)}(n,\vec{k})=\frac{\p E_n(k_\mu)}{\p k_\mu}=-\frac{\p_k f}{\p_E f}.
\end{align}   
Both derivatives in the right-hand side can be calculated analytically, so we should only insert the numerical solution for the $E_n(k)$ into the right-hand side of Eq.\eqref{eq:group-velocity}. In the case of magnetic field we are working in terms of $\tilde{\epsilon}=2\epsilon / \epsilon_0\equiv 2E/ \epsilon_0$. Then, we rewrite the group velocity as (and set $\hbar=1$) 
\begin{align}\label{eq:group-velocity-general}
	V_g^{(\mu)}(n,\vec{k})=\frac{\epsilon_0}{2}\frac{\p \tilde{\epsilon}_{n}(k)}{\p k_\mu} =-\frac{\epsilon_0}{2}\frac{\p_{k_\mu}f}{\p_{\tilde{\epsilon}}f}
\end{align} 
This is the most general expression, which we apply for all termination configurations. 

Now let us concentrate on a particular example of $C-C$ type termination. The corresponding spectral equation in $K$ valley is \cite{Bugaiko2019JPCM}
\begin{align}\label{eq:CC-characteristic}
	&U\left(-\frac{\widetilde{\epsilon}^2}{4} - \frac{\cos 2\Theta}{2},\sqrt{2}k_x l\right)V\left(-\frac{\widetilde{\epsilon}^2}{4} - \frac{\cos 2\Theta}{2},\sqrt{2}(k_x-k_0)l\right) - \nn
	&U\left(-\frac{\widetilde{\epsilon}^2}{4} - \frac{\cos 2\Theta}{2},\sqrt{2}(k_x-k_0)l\right)V\left(-\frac{\widetilde{\epsilon}^2}{4} - \frac{\cos 2\Theta}{2},\sqrt{2}k_x l\right) = 0,
\end{align}
where $k_0 = L/l^2$ is determined by the width of ribbon. The resulting spectrum is symmetric with respect to $k_0/2$, e.g. $\widetilde{\epsilon}(k_x)=\widetilde{\epsilon}(k_0-k_x)$. In panel (a) of Fig.\ref{fig:spectrum} we plot $\widetilde{\epsilon}(k_x)$ for the dice model case $\Theta=\frac{\pi}{4}$ with $L=5 l$. The spectrum form if qualitatively similar for all values of $\Theta\neq 0$. 

Using Eq.\eqref{eq:group-velocity-general} we found the group velocity along $x$ direction $V_{g}^{(x)}$ for arbitrary Landau level as a function of its index $n$, energy $\widetilde{\epsilon}$ and wave number $k_x$. The corresponding analytic expression is very complicated, so we present it in Eq.\eqref{eq:group-vel-cc}. Substituting numerically obtained solutions of Eq.\eqref{eq:CC-characteristic}, we plot the group velocity in panel (a) of Fig.\ref{fig:group-vel-zigzag}. The velocities demonstrate recurrent behavior near $k_x=k_0/2$ with growing index, namely, they start from zero value and the velocity for upper Landau level grows faster with wave number. Finally one should note that these results are qualitatively similar in both valleys, since for $\Theta\neq \frac{\pi}{4}$ the valley term adds only a constant energy shift to the whole dispersion $\tilde{\epsilon}_{n}(k_x)$. 

In the case of BA-AB termination the spectrum is also symmetric with respect to $k_0/2$. The corresponding spectral equation is presented in Appendix via \eqref{eq:AB_charact}. The peculiar property of such termination is that the dispersion of the first few Landau levels has a form of `Mexican hat' (the exact number of such levels depends on the ribbon size). This is manifested in the group velocity as large oscillations at $k_x$ near $k_0/2$ (see panel (b) in Fig.\ref{fig:group-vel-zigzag}). Also, for these levels the group velocity crosses zero three times, while for higher levels it crosses zero only in one point. 

A similar situation can be observed for BA-C boundary conditions, as shown on panel (c) in Fig.\ref{fig:group-vel-zigzag}. However, the spectrum in this case is not symmetric and the large oscillation in group velocity is present only at one side of the plot, for $k_x<k_0/2$. This can be understood by the fact that the wave number along the strip is linked with the quasi-classical center of motion for the electron orbit $k_{x}=y_{0} / R^{2}$, and here $R^{2}=\hbar c /|e| B$ is a cyclotron radius. Thus, panels (c) in Figs.\ref{fig:spectrum} and \ref{fig:group-vel-zigzag} demonstrate how the influences of each boundary type (BA or C) on electronic states interplay deep in the ribbon.
\begin{figure}
	\centering
	\hspace{-0.4cm}$\text{(a)}$\includegraphics[scale=0.45]{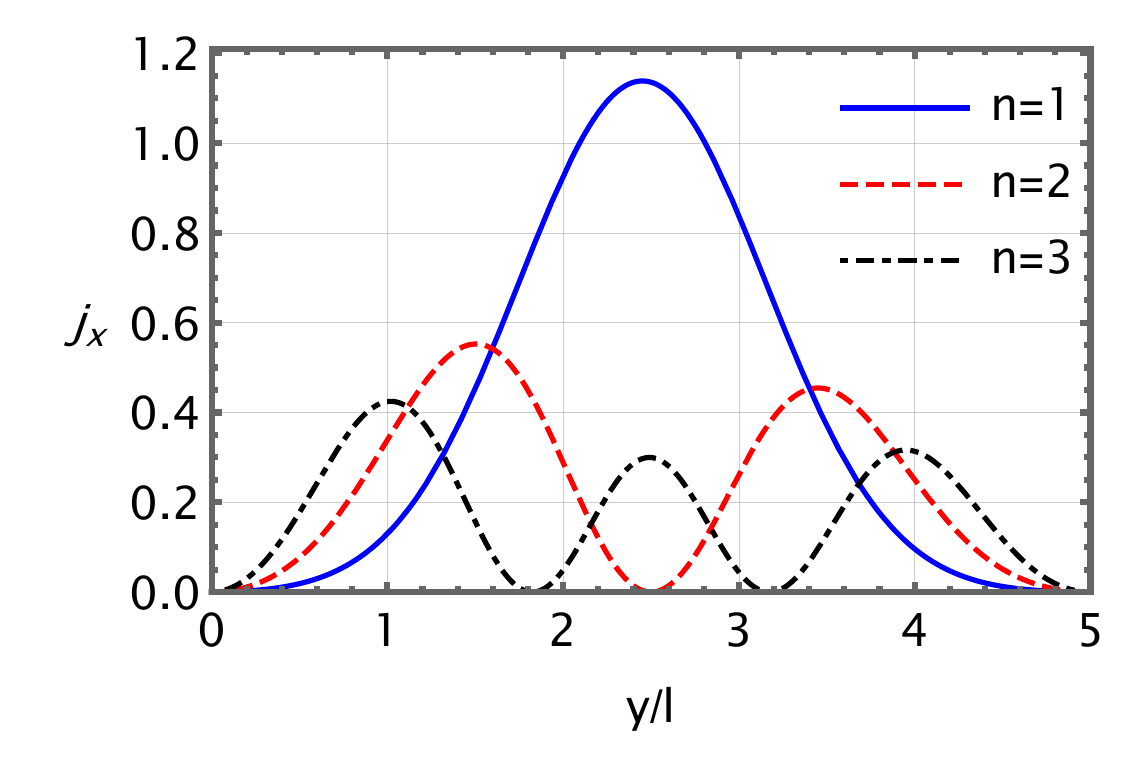}
	$\text{(b)}$\includegraphics[scale=0.45]{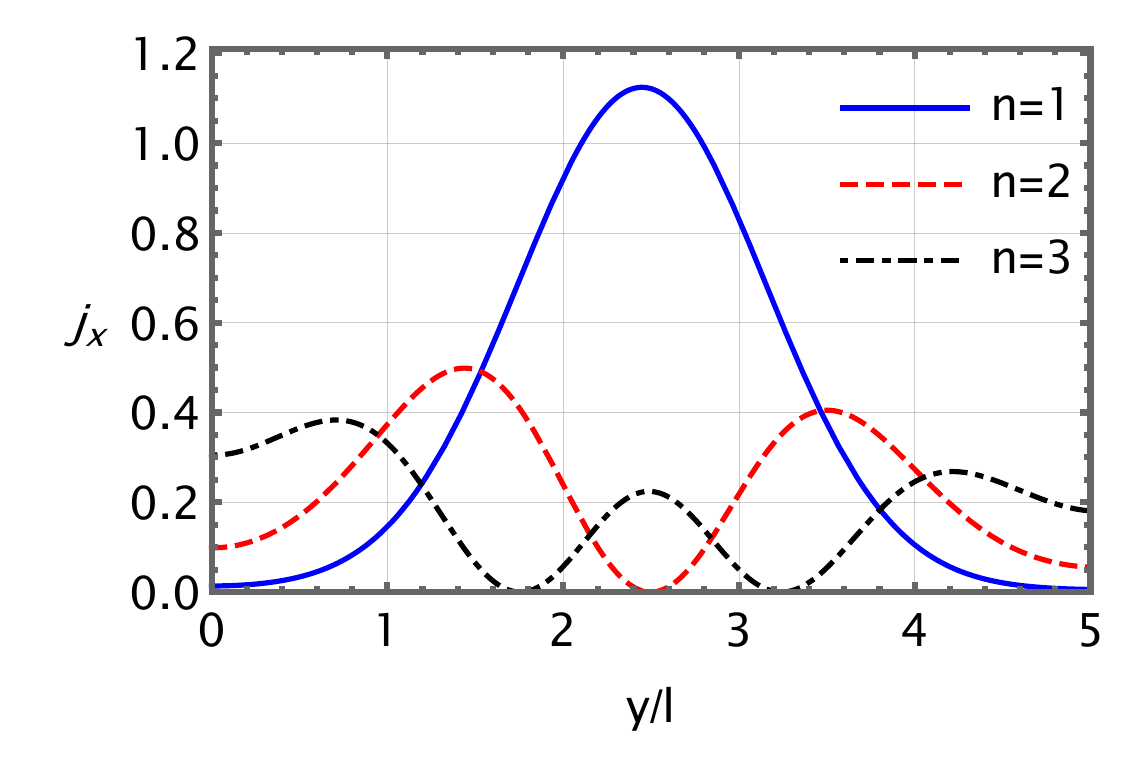}
	$\text{(c)}$\includegraphics[scale=0.45]{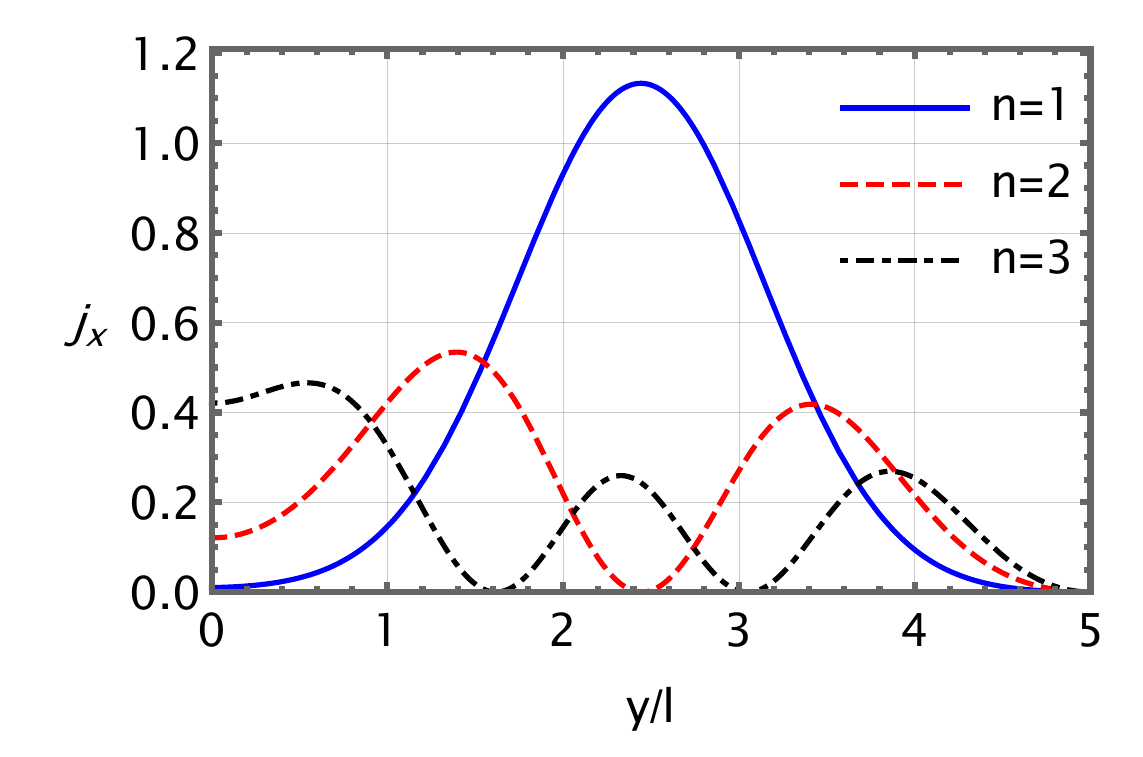}
	\caption{Currents distribution for zigzag terminated ribbon in the direction perpendicular to the edges of the ribbon. The parametric angle is $\Theta=\frac{\pi}{4}$ and we took the central wave number $k_x=k_0/2$. The size of the ribbon is $L=5 l$. Boundary condition types: panel (a) - C-C, panel (b) - BA-AB, panel (c) - BA-C. }
	\label{fig:currents-zigzag}
\end{figure}

The intuitive insight into the role of each boundary on group velocity can be obtain by analyzing the semi-infinite lattice with particular termination type. To find the edge states spectrum of such system, it is sufficient to take $L/l\to\infty$ limit in spectral equations \eqref{eq:CC-characteristic} or \eqref{eq:AB_charact}. The corresponding expressions for C-boundary and BA-boundary, which are accurate up to $k_x l^2$, have the following form \cite{Bugaiko2019JPCM}:
\begin{align}
	&\varepsilon_{C}(n,k_x, \Theta)=\pm \varepsilon_{0} \sqrt{2 n+\frac{3-\lambda \cos (2 \Theta)}{2}}\left(1-\frac{4 \Gamma(n+3 / 2)}{\pi n !(4 n+3-\lambda \cos (2 \Theta))} k_{x} l\right),\\
	&\varepsilon_{BA}(n,k_x, \Theta)=\pm \varepsilon_{0} \sqrt{2 n+\frac{1-\lambda \cos 2 \Theta}{2}}\left(1-\frac{\Gamma\left(n+\frac{1}{2}\right)}{\pi n !} k_{x} l\right)
\end{align}
 The corresponding group velocities are 
  \begin{align}
  	&V_{g,C}^{(x)}(n, \Theta)=\mp \frac{\varepsilon_{0} l}{\hbar} \frac{4 \Gamma(n+3 / 2) }{\sqrt{2}\pi n !\sqrt{4 n+3-\lambda \cos (2 \Theta)}} ,\quad V_{g,BA}^{(x)}(n, \Theta)=\mp \frac{\varepsilon_{0} l}{\hbar} \sqrt{2 n+\frac{1-\lambda \cos 2 \Theta}{2}}\frac{\Gamma\left(n+\frac{1}{2}\right)}{\pi n !}. 
  \end{align}
These expression describe the group velocities of edge states. Dividing the $V_{g,C}$ by $V_{g,BA}$, one can easily find that the $C$ edge group velocity is always larger than $BA$ one in the $K$ valley for $\Theta\leq \frac{\pi}{4}$. This relation is not very universal, since these expressions do not take into account next powers in $k_x l$. Also, one can these expressions show that $V_{g,C}$ decays with index $n$ and $V_{g,BA}$ grows with $n$ for $\Theta<\frac{\pi}{4}$.

\subsection{Current distribution}
Next, let us proceed to the analysis of currents distribution. 
The electric current in the $x$ direction, which is defined as $\la\Psi|\lambda S_x|\Psi\ra$ for the state $\Psi$ with definite $k_x$ and $\tilde{\epsilon}$, has the following expression through sublattice components \cite{Bugaiko2019JPCM}:
\begin{align}\label{eq:current-zigzag}
	j_{x}(k_x, \tilde{\epsilon})&=\lambda\bigg[\cos\Theta\, \psi_A^{*}\psi_C+\sin\Theta\, \psi_B^{*}\psi_C+h.c.\bigg]=\frac{\sqrt{2} \lambda}{\tilde{\varepsilon}}\left[\lambda \xi \psi_{C} \psi_{C}^{*}+\cos 2 \Theta \psi_{C} \partial_{\xi} \psi_{C}^{*}+h . c .\right].
\end{align}
In the second line we used the expressions for $\psi_A$ and $\psi_B$ found from Schrodinger equation. 
To evaluate this expression we need to find the exact normalized solutions in each valley. Thus, to determine the constants $C_1$ and $C_2$ in general solution \eqref{eq:zigzag_general} we use the boundary condition at $y=0$. We find:
\begin{align}\label{eq:C_2-C1-relation}
	C_2=-C_1 U\left(-\frac{\widetilde{\varepsilon}^{2}}{4}-\frac{\cos 2 \Theta}{2}, \sqrt{2} k_{x} l\right) \bigg/\bigg. V\left(-\frac{\widetilde{\varepsilon}^{2}}{4}-\frac{\cos 2 \Theta}{2}, \sqrt{2} k_{x} l\right).
\end{align}
Note that the equation at $y=L$ gives the same relation for the spectral solutions $\tilde{\epsilon}(k_x)$. Next, to determine $C_1$ we apply normalization condition 	$\int_{0}^{L}dy\left(\sum{}_{i=A,B,C}\psi_{i}^{*}\psi_i\right)=1$, which holds true for each valley separately \cite{Oriekhov2018FNT}.
Performing the integration numerically, and substituting $C_{1,2}$ into Eq.\eqref{eq:current-zigzag}, we find the current distribution. We plot this distribution for several lowest Landau levels on Fig.\ref{fig:currents-zigzag}, taking $k_x=k_0/2$. 

One should note that the current distribution for $n=1$ level is nearly the same for all three configurations C-C, BA-AB and BA-C. However, near the edge with C-termination the current tends to exactly zero value. This holds true for all Landau levels, since the total expression for current \eqref{eq:current-zigzag} is proportional to $\psi_C$ component, which is zero at the boundary. Near the BA boundary the current is nonzero and grows with Landau level index $n$. This suggests

\section{Ribbon infinite in y-direction: armchair termination}
\label{sec:armchair}
\begin{figure*}
	\centering
	$\text{(a)}$\includegraphics[scale=0.43]{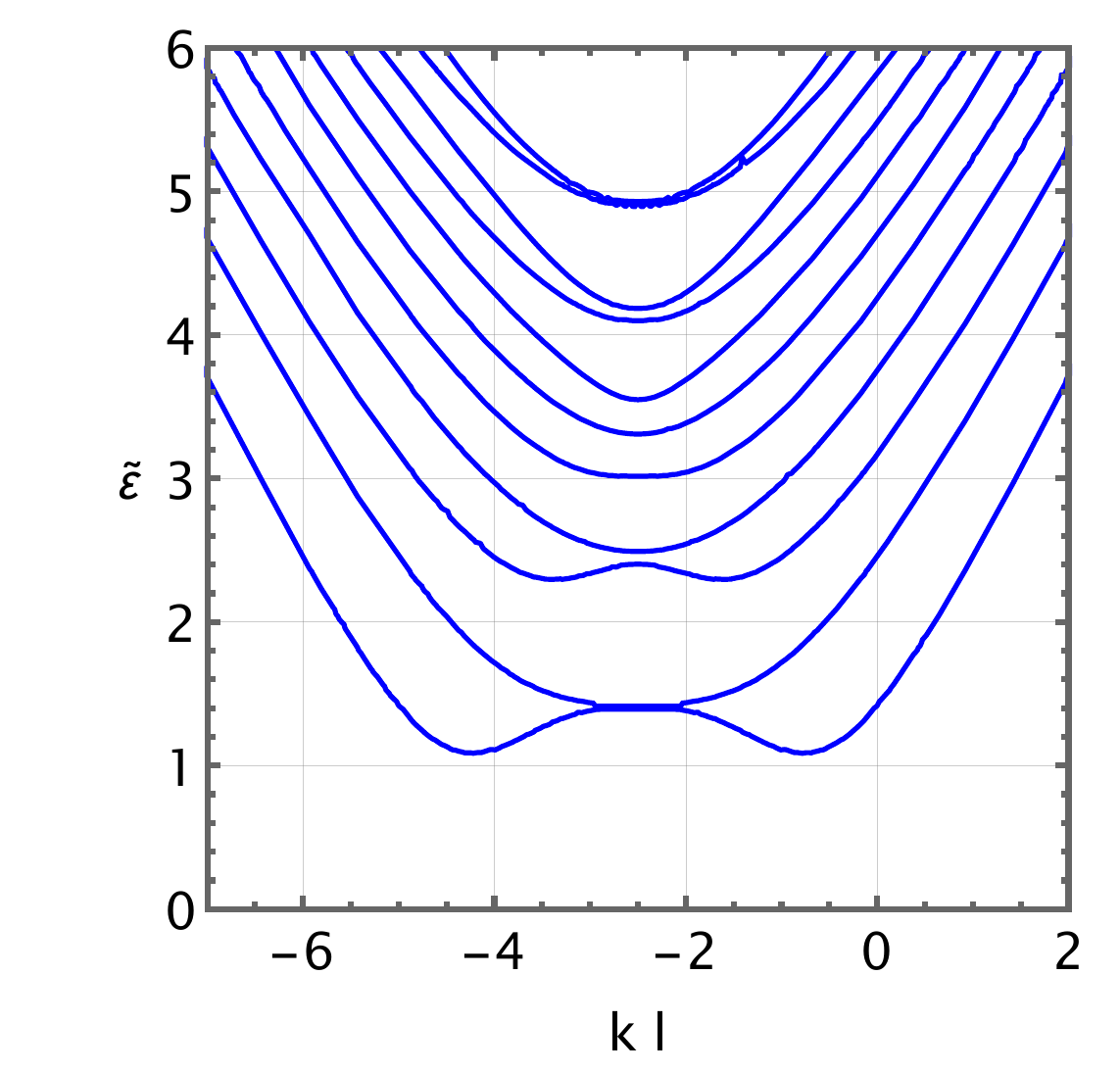}\qquad\qquad
	$\text{(b)}$\includegraphics[scale=0.43]{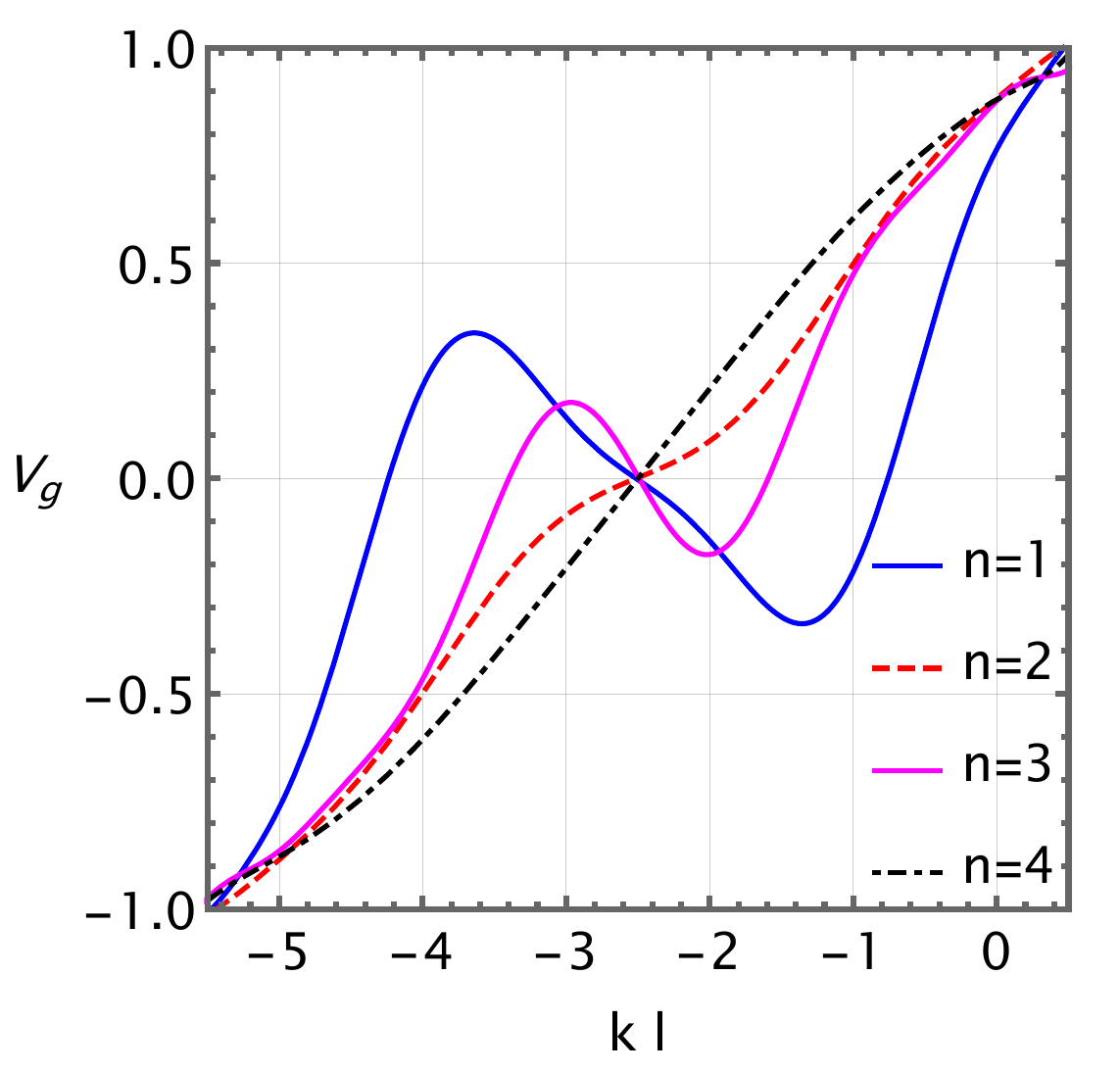}
	\caption{Panel (a): spectrum for the ribbon with armchair termination, for $L=5 l$, and $\cos\Delta KL=1$. Panel (b) - group velocity $V_g$ for the lowest four Landau levels. One shiuld note the presence of strong oscillations in $V_g$ for $n=1$ and $n=3$ levels, that are associated with "Mexican hat" shape of corresponding energy levels.}
	\label{fig:arm_spectr_vel}
\end{figure*}
The ribbon with armchair termination has a translational invariance along $y$ direction (see Fig.\ref{fig1}, panel (c)). Thus, it is convenient to
use the vector potential in the following gauge: $A = (0, Bx)$. The wave functions can be taken in the form  $\Psi=e^{ik_y y}\psi(x)$, with $\psi(x)$ defined by the following Schrodinger equation \cite{Bugaiko2019JPCM}
\begin{align}\label{eq:hamiltonian_armchair}
	i
	\begin{pmatrix}
		0 & \cos\Theta(-\lambda\partial_\xi - \xi) & 0\\
		\cos\Theta(-\lambda\partial_\xi + \xi) & 0 & \sin\Theta(-\lambda\partial_\xi - \xi)\\
		0& \sin\Theta(-\lambda\partial_\xi + \xi) & 0
	\end{pmatrix}
	\begin{pmatrix}
		\psi_A \\ \psi_C \\ \psi_B
	\end{pmatrix}
	=
	\frac{\widetilde{\epsilon}}{\sqrt{2}}
	\begin{pmatrix}
		\psi_A \\ \psi_C \\ \psi_B
	\end{pmatrix}.
\end{align}
In this equation we defined the variable $\xi = k_y l + x/l$ (note the plus sign before $x$). In each valley this system reduces to the  second-order equation \eqref{eq:eq_on_psi_C} for the $\phi_C$ component. The armchair boundary condition at $x=0$ and $x=L$ edges implies \cite{Oriekhov2018FNT,Akhmerov,Bugaiko2019JPCM}:
\begin{align}\label{eq:armchair_conditions}
	\psi_{\mu}(x=0)=\psi_{\mu^{\prime}}(x=0),\quad \psi_{\mu}(x = L) = e^{i \Delta KL}\psi_{\mu^{\prime}}(x = L).
\end{align}
Here $\mu=A,\,B,\,C$ denoted the sublattice index. The second boundary condition contains a phase factor $e^{i \Delta KL}$ that depends distance between K and K' points in momentum space in $k_x$ direction, $\left(\mathbf{K}-\mathbf{K}^{\prime}\right)\left(L \mathbf{e}_{x}\right)=\Delta K L=4 \pi L/ 3 \sqrt{3} d$. The spectral equation for these boundary conditions is presented in Appendix, see Eq.\eqref{eq:armchair-spec}. 
From this equation we evaluate the group velocity for 4 lowest Landau levels by using the general relation \eqref{eq:group-velocity-general}. Since the spectrum in perpendicular  magnetic field practically does not change with the $\Delta KL$ value \cite{Bugaiko2019JPCM}. This is because all levels become gapped, and the qualitative difference between $\cos\Delta KL=1$ and $\cos\Delta KL=-1/2$, that was noted without magnetic field \cite{Oriekhov2018FNT} disappears.
Thus, we plot the spectrum and group velocity in Fig.\ref{fig:arm_spectr_vel} only for the case $\cos\Delta KL=1$. The `Mexican hat` shape of the spectrum for the few lowest Landau levels with odd index ($n=1$ and $n=3$ in our case) is manifested as a large oscillation in group velocity profile.   

\begin{figure}
	\centering
	$\text{(a)}$\includegraphics[scale=0.45]{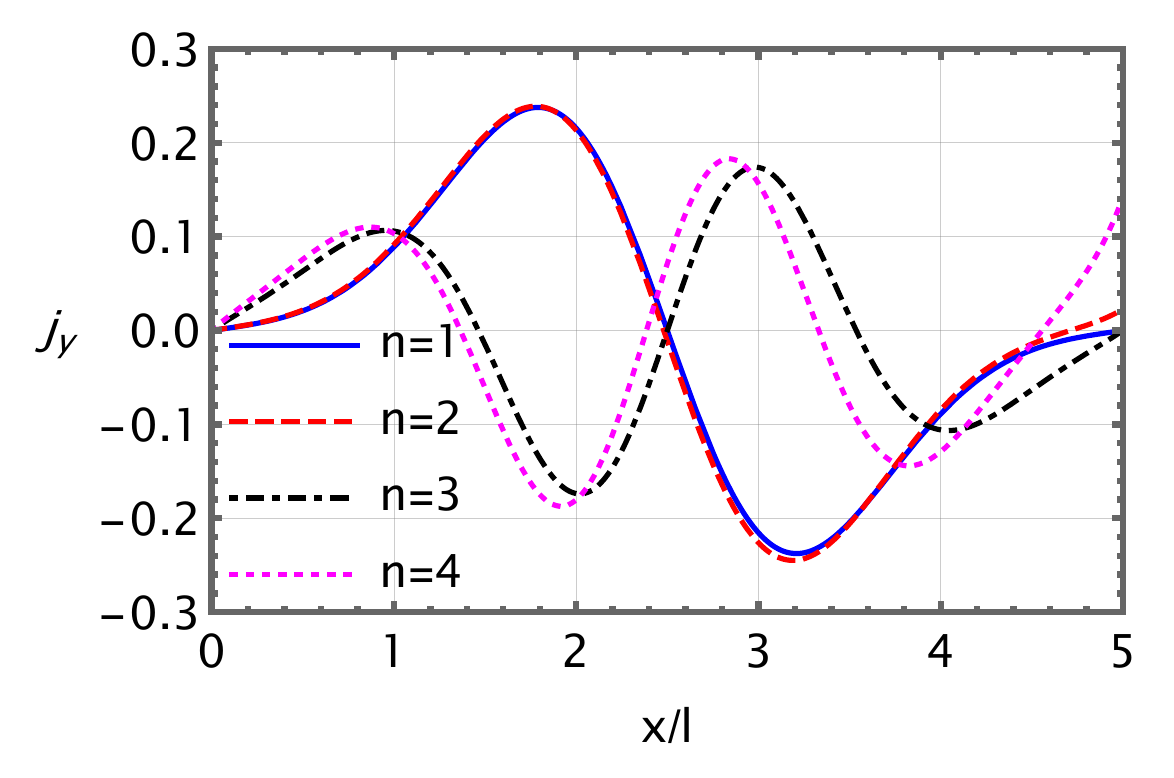}
	$\text{(b)}$\includegraphics[scale=0.45]{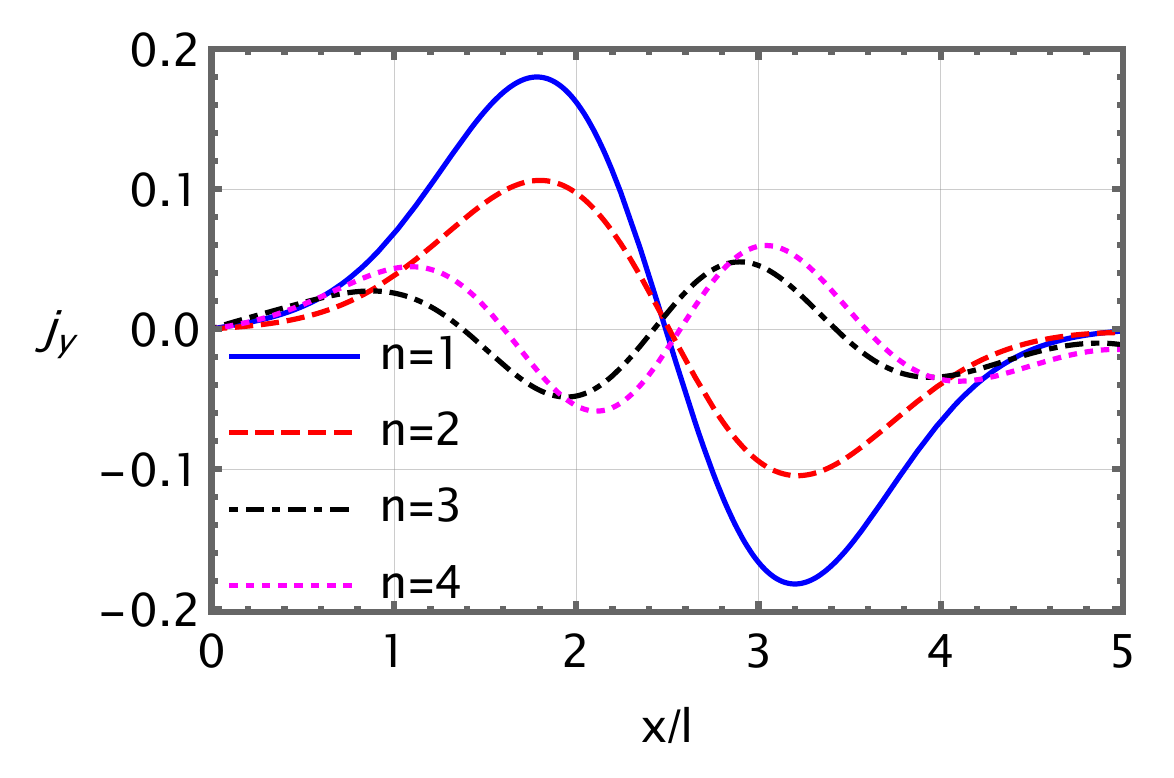}
	$\text{(b)}$\includegraphics[scale=0.45]{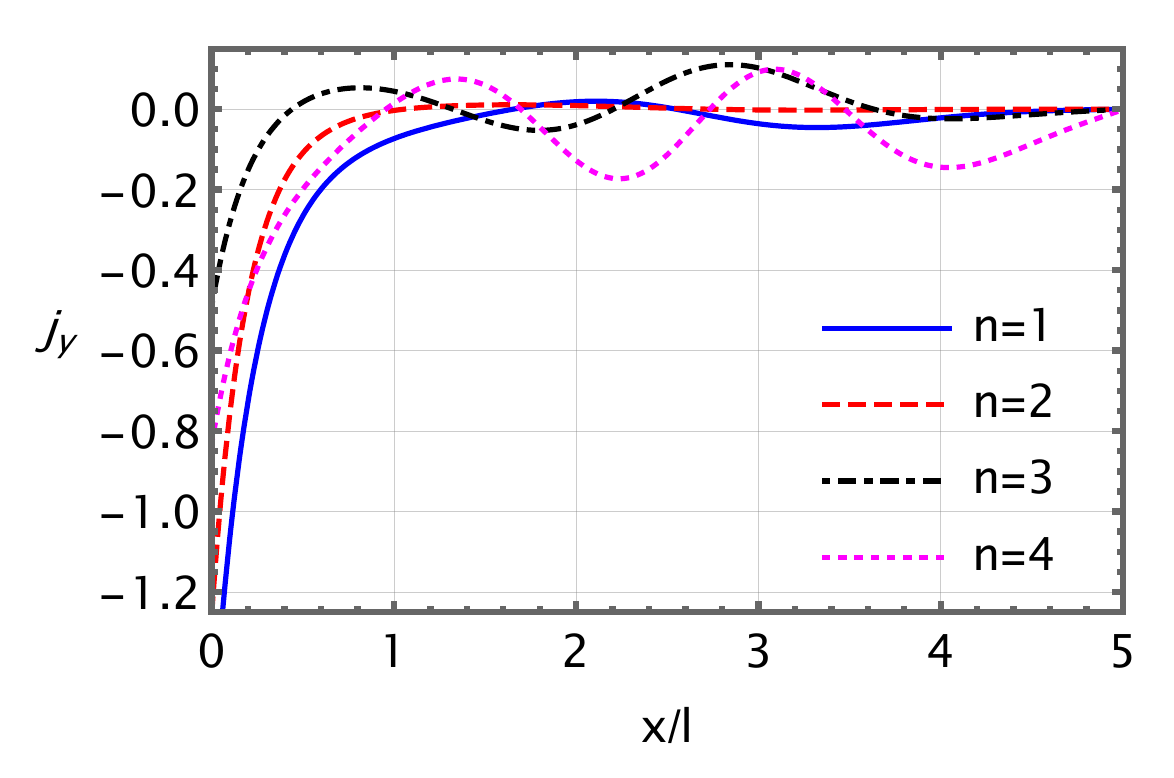}
	\caption{Currents distribution for armchair terminated ribbon in the direction perpendicular to the edges of the ribbon for the lowest Landau levels with nonzero energy. The parametric angle is $\Theta=\frac{\pi}{4}$ for panels (a) and (b) and we took the central wave number $k_y=-k_0/2$. The size of the ribbon is $L=5 l$. On the panel (c) we took $\Theta=\frac{\pi}{5}$. The panels (a) and (b) present the cases with $\cos\Delta KL=1$ and $-1/2$ respectively.}
	\label{fig:arm-current}
\end{figure}

Next we proceed to the evaluation of current. Is is given by the following expression:
\begin{align}
	j_{y}(k_y, \tilde{\epsilon})=\bigg[-i\cos\Theta\, \psi_A^{*}\psi_C+i\sin\Theta\, \psi_B^{*}\psi_C+h.c.\bigg]+(\lambda\to-\lambda),
\end{align}
and now contains the terms from each valley. Substituting the expressions for $\psi_A$ and $\psi_B$ in terms of $\psi_C$, we find
\begin{align}\label{eq:arm-current}
	j_{y}(k_y, \tilde{\epsilon})=\frac{\sqrt{2}}{\tilde{\epsilon}}\left(\cos2\Theta \lambda\xi \Psi_C^{*}(\lambda,\dots)\Psi_C(\lambda,\dots)+\Psi_C\p_\xi \Psi_C^{*}\right)+(\lambda\to-\lambda).
\end{align}
 We find that difference between the two possible $\cos\Delta KL$ values is crucial for the current distribution profile as a function of coordinate $x$. This is due to the fact that the phase factor $e^{i\Delta K L}$ also appears in the system of equations for $C_{1,2}$ and $C_{1,2}^{'}$ constants, see Eq.\eqref{eq:arm-C-const} in Appendix. The two different cases for $\cos\Delta KL=1$ (panel (a)) and $\cos\Delta KL=-1/2$ (panel (b)) are plotted on Fig.\ref{fig:arm-current}. We took he relation of ribbon width to magnetic length to be the same in both cases, $L/l=5$, and plotted the current for the central wave number $k_y=-L/2l^2$. This works well when both $L$ and $l$ are much larger that the lattice constant $a$. We used parametric angle $\Theta=\frac{\pi}{4}$ for panel (a) and (b) that corresponds to the dice model, and $\Theta=\frac{\pi}{5}$ for (c) panel. 
 
 One should point out several main differences between three distributions, plotted in Fig.\ref{fig:arm-current}. On panel (a) ($\cos\Delta KL=1$) the number of points at which $j_{y}(x)$ crosses zero coincides with the Landau level index. Also, the states with even index have nonzero current on $x=L$ boundary, which suggests about the formation of current-carrying edge states. In the bulk the oscillations of current have nearly the same amplitude. At the same time, on the panel (b)($\cos\Delta KL=-1/2$) the amplitude of oscillations drastically reduces for $n=2$ level comparing to $n=1$. Also, the current is approximately zero for all levels at $x=L$ edge. One should note that the current is zero at $x=0$ edge only in dice model, for which the term with $\cos(2\Theta)$ in Eq.\eqref{eq:arm-current} vanishes. In the case of $\Theta\neq\frac{\pi}{5}$ the current reaches its maximum at the edge $x=0$ for all studied levels. This is qualitatively similar to graphene case, discussed in Ref.\cite{Wang2011}. At small angles the $\alpha-\mathcal{T}_3$ model becomes similar to graphene, despite the presence of third, completely flat band. However, this flat band does not have nonzero group velocity and does not carry any current, since it consists of localized states.

\section{Conclusions}
\label{sec:conclusion}
In the present paper we studied the group velocities and distributions of current in the ribbons made of $\mathcal{T}_3$ lattice placed in perpendicular magnetic field. Using effective low-energy model, we performed the analysis for all combinations of simple boundary conditions of zigzag and armchair type. It is important to note that the flat band does not play any role in current distribution since it consists of localized states. On algebraic level it is manifested through the fact that the current is always proportional to the C-component of wave function, which is zero for flat band solutions also in a magnetic field \cite{Bercioux,Bugaiko2019JPCM}.

In this paper we concentrated attention mainly on zigzag-terminated ribbons. For the $\alpha-\mathcal{T}_3$ model zigzag boundary conditions demonstrate much larger variety of regimes than in graphene.  Particularly, we found that the formation of edge current is possible near the $BA$-type boundary. For the $C$-type boundary the current is always zero, because it is proportional to the $C$-component of wave function. Notably, the current is always positively defined and the number of oscillations per ribbon equals the the index of Landau level. Also we discussed the semi-infinite lattice and found the exact role of each boundary type on group velocity, which is constant (larger near C-boundary) near the edge.
 
In the case of armchair terminated ribbon we found, that while there is no qualitative difference in the spectrum for different ribbon width, the current distribution is strongly influenced by width type. For the ribbon with "metallic" width ($L=(\sqrt{3} / 2)(\tilde{N}+1) d$, with $\tilde{N}=3 N-1$ is the number of atomic rows \cite{Oriekhov2018FNT}) the formation of current-carrying edge states is possible. Notably, such edge states are linked with the Landau levels with even index. In the opposite case of "insulating" width the edge currents are not observed for the case of dice model. At parametric angles $\Theta< \frac{\pi}{4}$ the armchair ribbons demonstrate very similar behavior to graphene ribbons, for which the current has strong peak near one of the edges \cite{Wang2011}. Also, the current has alternating sign inside the ribbon. The current distributions reach their maximum values in the bulk in dice model $\Theta=\frac{\pi}{4}$ and on the edge for $\Theta<\frac{\pi}{4}$. This can be linked to the fact that for the smaller angles $\Theta<\frac{\pi}{4}$ the model is more similar to graphene with weakly coupled additional sites inside hexagons. 

Finally, we note that these results can be important to further investigate the formation and properties of edge states near each boundary type for terminated dice lattice. Another open question is the formation of edge gapless states in the ribbons with gapped $\alpha-\mathcal{T}_3$ model. Refs.\cite{Xu2017PRB,Dey2020PRB-Haldane} considered dice ribbons with $S_z$ gap term in the low energy model and found such states. However, the formation of such states was not considered for the intervalley gap term \cite{Gorbar2021} or the $\Delta \text{diag}(1,-1,1)$ gap \cite{Raoux2015JPCM}. The intervalley gap (which couples states from different valleys) was introduced recently in Ref.\cite{Gorbar2021}, and it was shown that dynamical generation of this gap is strongly enhanced comparing to gap terms in one valley.

\begin{acknowledgments}
	We are grateful to E. V. Gorbar and V. P. Gusynin for useful discussions and critical reading of the manuscript.
\end{acknowledgments}

\appendix
\section{Dispersion relations for zigzag- and armchair-terminated ribbons}
\label{appendix}
In this Appendix we recall the dispersion relations for zigzag-type terminated ribbons, which were derived in Ref.\cite{Bugaiko2019JPCM}. Also, we present some results for the group velocities. 

\subsection{Zigzag termination}

The spectrum plots for all three types of termination combinations $C-C$, $AB-BA$ and $AB-C$ are presented on Fig.\ref{fig:spectrum}. 

In the C-C case the group velocity has the following form:
The group velocity is given by 
\begin{align}\label{eq:group-vel-cc}
	&v_x(k_x,\tilde{\epsilon})=-\frac{\epsilon_0}{\tilde{\epsilon}}\times\nn
	&\left[2^{a+\frac{1}{2}} l e^{-\frac{1}{2} l^2
		\left(\left(k_0-k_x\right){}^2+k_x^2\right)}
	\left(H_{-a-\frac{1}{2}}\left(l k_x\right) H_{\frac{1}{2}-a}\left(l
	\left(k_0-k_x\right)\right)+H_{-a-\frac{1}{2}}\left(-l k_x\right)
	H_{\frac{1}{2}-a}\left(l
	\left(k_x-k_0\right)\right)+\right.\right.\nn
	&+\left.\left.	H_{-a-\frac{1}{2}}\left(l
	\left(k_x-k_0\right)\right) \left(l \left(k_0-2 k_x\right)
	H_{-a-\frac{1}{2}}\left(-l k_x\right)-H_{\frac{1}{2}-a}\left(-l
	k_x\right)\right)-\right.\right.\nn
	&\left.\left. -H_{-a-\frac{1}{2}}\left(l
	\left(k_0-k_x\right)\right) \left(H_{\frac{1}{2}-a}\left(l
	k_x\right)+l \left(k_0-2 k_x\right) H_{-a-\frac{1}{2}}\left(l
	k_x\right)\right)\right)\right]\times\nn
	&\times\left[U(a, \sqrt{2} l
	k_x)
	D^{(1,0)}_{-a-\frac{1}{2}}\left(\sqrt{2} l
	\left(k_0-k_x\right)\right)-D_{-a-\frac{1}{2}}\left(-\sqrt{2} l
	k_x\right)
	D^{(1,0)}_{-a-\frac{1}{2}}\left(\sqrt{2} l
	\left(k_x-k_0\right)\right)+\right.\nn
	&\left.+U(a, \sqrt{2} l
	\left(k_x-k_0\right)) \left(\psi
	^{(0)}\left(a+\frac{1}{2}\right) D_{-a-\frac{1}{2}}\left(-\sqrt{2} l
	k_x\right)-D^{(1,0)}_{-a-\frac{1}{2}}\left(-\sqrt{2} l k_x\right)\right)+\right.\nn
	&\left.+D_{-a-\frac{1}{2}}\left(\sqrt{2} l
	\left(k_0-k_x\right)\right)
	\left(D^{(1,0)}_{-a-\frac{1}{2}}\left(\sqrt{2}
	l k_x\right)-\psi ^{(0)}\left(a+\frac{1}{2}\right)
	D_{-a-\frac{1}{2}}\left(\sqrt{2} l k_x\right)\right)\right]^{-1}
\end{align}
Here $H_{a}(x)$ are Hermite polynomials, $D_{a}(x)$ are the parabolic cylinder functions (see \cite{Abramowitz} and appendix in Ref.\cite{Bugaiko2019JPCM} for the relation between $D$ functions and $U,\,V$) and $\psi^{0}(x)$ is the polygamma function. This expression is relatively complicated. We can obtain similar expressions in all other cases, but they are even more cumbersome. Thus, we choose to plot these group velocities for numerically obtained solutions $\tilde{\epsilon}(k_x)$ in Fig.\ref{fig:group-vel-zigzag}.

\subsection{The \textbf{BA}-\textbf{AB} zigzag termination}
In case of AB-AB termination we have the following characteristic equation
\begin{align}\label{eq:AB_charact}
	&\hspace{-0.5em}\left[\sqrt{2} U^{\prime}\left(a, \sqrt{2} k_{x} l\right)+\cos 2 \Theta k_{x} l U\left(a, \sqrt{2} k_{x} l\right)\right]\left[\sqrt{2} V^{\prime}\left(a,\sqrt{2}\left(k_{x}-k_{0}\right) l\right)+\cos 2 \Theta\left(k_{x}-k_{0}\right) l V\left(a,\sqrt{2}\left(k_{x}-k_{0}\right) l\right)\right]-  \nn
	&\hspace{-0.5em}\left[\sqrt{2} U^{\prime}\left(a,\sqrt{2}\left(k_{x}-k_{0}\right) l\right)+\cos 2 \Theta\left(k_{x}-k_{0}\right) l U\left(a,\sqrt{2}\left(k_{x}-k_{0}\right) l\right)\right]\left[\sqrt{2} V^{\prime}\left(a,\sqrt{2} k_{x} l\right)+\cos 2 \Theta  k_{x} l V\left(a,\sqrt{2} k_{x} l\right)\right]=0.
\end{align}

To evaluate the current, we use the following solution for the $C_2$ constant in terms of $C_1$
\begin{align}
	C_2 =-C_1\left[\sqrt{2} U^{\prime}\left(a, \sqrt{2} k_{x} l\right)+\cos 2 \Theta k_{x} l U\left(a, \sqrt{2} k_{x} l\right)\right]\left[\sqrt{2} V^{\prime}\left(a,\sqrt{2} k_{x} l\right)+\cos 2 \Theta  k_{x} l V\left(a,\sqrt{2} k_{x} l\right)\right]^{-1}.
\end{align}

\subsection{The \textbf{C}-\textbf{AB} zigzag termination}
The characteristic equation has the form
\begin{align}
	&\left[\sqrt{2} U^{\prime}\left(a, \sqrt{2} k_{x} l\right)+\cos 2 \Theta k_{x} l U\left(a, \sqrt{2} k_{x} l\right)\right]V\left(-\frac{\widetilde{\epsilon}^2}{4} - \frac{\cos 2\Theta}{2},\sqrt{2}(k_x-k_0)l\right)-\nn
	&U\left(-\frac{\widetilde{\epsilon}^2}{4} - \frac{\cos 2\Theta}{2},\sqrt{2}(k_x-k_0)l\right) \left[\sqrt{2} V^{\prime}\left(a,\sqrt{2} k_{x} l\right)+\cos 2 \Theta  k_{x} l V\left(a,\sqrt{2} k_{x} l\right)\right]=0.
\end{align}
To evaluate the current, we use the following solution for the $C_2$ constant in terms of $C_1$
\begin{align}
	C_2=-C_1 U\left(-\frac{\widetilde{\epsilon}^2}{4} - \frac{\cos 2\Theta}{2},\sqrt{2}(k_x-k_0)l\right)\left[V\left(-\frac{\widetilde{\epsilon}^2}{4} - \frac{\cos 2\Theta}{2},\sqrt{2}(k_x-k_0)l\right)\right]^{-1}.
\end{align}
which is the same as in C-C termination case (compare with Eq.\eqref{eq:C_2-C1-relation}), despite the fact that substituted energy $\tilde{\epsilon}(k_x)$ is different.

\subsection{Armchair termination}
The set of armchair boundary conditions can be rewritten as 
\begin{align}
		\psi_{C}^{\prime}=-\left.\psi_{C^{\prime}}^{\prime}\right|_{x=0}, \quad \psi_{C}=\left.\psi_{C^{\prime}}\right|_{x=0}, \quad \psi_{C^{\prime}}^{\prime}=-\left.\mathrm{e}^{\mathrm{i} \Delta K L} \psi_{C^{\prime}}^{\prime}\right|_{x=L},\quad 
		\psi_{C}=\left.\mathrm{e}^{\mathrm{i} \Delta K L} \psi_{C^{\prime}}\right|_{x=L}.
\end{align}
Substituting the solution for $\psi_{C}$ in each valley, we find the following system of equations for the free constants $C_{1,2}$ and $C_{1,2}^{'}$:
\begin{align}\label{eq:arm-C-const}
	\left(
	\begin{array}{cccc}
		U\left(\epsilon_1,\xi_1\right) &
		V\left(\epsilon_1,\xi_1\right) &
		-U\left(\epsilon_2,\xi_1\right) &
		-V\left(\epsilon_2,\xi_1\right) \\
		U\left(\epsilon_1,\xi_2\right) &
		V\left(\epsilon_1,\xi_2\right) & -e^{i \Delta K L }
		U\left(\epsilon_2,\xi_2\right) & -e^{i \Delta K L  }
		V\left(\epsilon_2,\xi_2\right) \\
		\frac{\sqrt{2}}{l} U'\left(\epsilon_1,\xi_1\right) &
		\frac{\sqrt{2}}{l} V'\left(\epsilon_1,\xi_1\right) &
		\frac{\sqrt{2}}{l} U'\left(\epsilon_2,\xi_1\right) &
		\frac{\sqrt{2}}{l} V'\left(\epsilon_2,\xi_1\right) \\
		\frac{\sqrt{2}}{l} U'\left(\epsilon_1,\xi_2\right) &
	\frac{\sqrt{2}}{l} V'\left(\epsilon_1,\xi_2\right) &
		e^{i \Delta K L  } \frac{\sqrt{2}}{l} U'\left(\epsilon_2,\xi_2\right) & e^{i \Delta K L  } \frac{\sqrt{2}}{l}
		V'\left(\epsilon_2,\xi_2\right) \\
	\end{array}
	\right)\begin{pmatrix}
		C_1\\
		C_2\\
		C_1^{'}\\
		C_{2}^{'}
	\end{pmatrix}=0.
\end{align}
where $\epsilon_1 = -\frac{\widetilde{\epsilon}^2}{4} + \frac{\cos2\Theta}{2}$,
$\epsilon_2 = -\frac{\widetilde{\epsilon}^2}{4} - \frac{\cos2\Theta}{2}$, $\xi_1 = \sqrt{2}k_yl$, $\xi_2 = \sqrt{2}(k_y l + L/l)$.

The dispersion relation for the ribbon with armchair edges is a solution of the following equation:
\begin{align}\label{eq:armchair-spec}
	\frac{4}{\pi} \cos \Delta K L&-\left(U^{\prime}\left(\varepsilon_{1}, \xi_{1}\right) V\left(\varepsilon_{2}, \xi_{1}\right)+V^{\prime}\left(\varepsilon_{2}, \xi_{1}\right) U\left(\varepsilon_{1}, \xi_{1}\right)\right)\left(U\left(\varepsilon_{2}, \xi_{2}\right) V^{\prime}\left(\varepsilon_{1}, \xi_{2}\right)+V\left(\varepsilon_{1}, \xi_{2}\right) U^{\prime}\left(\varepsilon_{2}, \xi_{2}\right)\right)\nn
	&+\left(V^{\prime}\left(\varepsilon_{1}, \xi_{1}\right) V^{\prime}\left(\varepsilon_{2}, \xi_{1}\right)+V^{\prime}\left(\varepsilon_{1}, \xi_{1}\right) V\left(\varepsilon_{2}, \xi_{1}\right)\right)\left(U\left(\varepsilon_{1}, \xi_{2}\right) U^{\prime}\left(\varepsilon_{2}, \xi_{2}\right)+U\left(\varepsilon_{2}, \xi_{2}\right) U^{\prime}\left(\varepsilon_{1}, \xi_{2}\right)\right)\nn
	&+\left(U\left(\varepsilon_{1}, \xi_{1}\right) U^{\prime}\left(\varepsilon_{2}, \xi_{1}\right)+U^{\prime}\left(\varepsilon_{1}, \xi_{1}\right) U\left(\varepsilon_{2}, \xi_{1}\right)\right)\left(V\left(\varepsilon_{2}, \xi_{2}\right) V^{\prime}\left(\varepsilon_{1}, \xi_{2}\right)+V^{\prime}\left(\varepsilon_{2}, \xi_{2}\right) V\left(\varepsilon_{1}, \xi_{2}\right)\right)\nn
	&-\left(V\left(\varepsilon_{1}, \xi_{1}\right) U^{\prime}\left(\varepsilon_{2}, \xi_{1}\right)+V^{\prime}\left(\varepsilon_{1}, \xi_{1}\right) U\left(\varepsilon_{2}, \xi_{1}\right)\right)\left(U\left(\varepsilon_{1}, \xi_{2}\right) V^{\prime}\left(\varepsilon_{2}, \xi_{2}\right)+U^{\prime}\left(\varepsilon_{1}, \xi_{2}\right) V\left(\varepsilon_{2}, \xi_{2}\right)\right)=0
\end{align}
where the coefficient near $\cos\Delta KL$, is the Wronskian of parabolic cylinder functions $\mathcal{W}=\sqrt{2/\pi}$ \cite{Abramowitz}.

\end{document}